\documentstyle[preprint,bookman,aps]{revtex}
\def\singlespace {\smallskipamount=3.75pt plus1pt minus1pt
                  \medskipamount=7.5pt plus2pt minus2pt
                  \bigskipamount=15pt plus4pt minus4pt
                  \normalbaselineskip=15pt plus0pt minus0pt
                  \normallineskip=1pt
                  \normallineskiplimit=0pt
                  \jot=3.75pt
                  {\def\smallskip {\vskip\smallskipamount}}
                  {\def\medskip   {\vskip\medskipamount}}
                  {\def\bigskip   {\vskip\bigskipamount}}
                  {\setbox\strutbox=\hbox{\vrule
                    height10.5pt depth4.5pt width 0pt}}
                  \parskip 4.5pt
                  \normalbaselines}
\def\middlespace {\smallskipamount=5.825pt plus1.5pt minus1.5pt
                  \medskipamount=11.25pt plus3pt minus3pt
                  \bigskipamount=22.5pt plus6pt minus6pt
                  \normalbaselineskip=22.5pt plus0pt minus0pt
                  \normallineskip=1pt
                  \normallineskiplimit=0pt
                  \jot=5.825pt
                  {\def\smallskip {\vskip\smallskipamount}}
                  {\def\medskip   {\vskip\medskipamount}}
                  {\def\bigskip   {\vskip\bigskipamount}}
                  {\setbox\strutbox=\hbox{\vrule
                    height15.75pt depth6.75pt width 0pt}}
                  \parskip 7.25pt
                  \normalbaselines}

\def\be{\begin{equation}}
\def\lan{\left\langle}
\def\ran{\right\rangle}
\def\ee{\end{equation}}
\def\barr{\begin{array}}
\def\earr{\end{array}}

\def\nn8{\nonumber\\[10pt]}
\def\l{\left}
\def\r{\right}
\def\dis{\displaystyle}
\def\ed{\end{document}}
\def\spin{\frac{1}{2}}

\begin{document}

\title{Classification of States in $O(8)$ Proton-Neutron \\ Pairing  Model}

\author{V. K. B.  KOTA\footnote{Fax: 91-79-26301502. {\it e-mail address}:
vkbkota@prl.ernet.in (V.K.B. Kota)}}
\address{Physical Research Laboratory, Ahmedabad 380 009, India}
\author{J. A. CASTILHO ALCAR\'AS\footnote{{\it e-mail address}:
jaca@ift.unesp.br (J. A. Castilho Alcaras)}}
\address{Instituto de F\'{\i}sica Te\'orica, UNESP, 01405-900, 
S\~{a}o Paulo, Brazil}

\maketitle

\begin{abstract}

Isoscalar ($T=0$) plus isovector ($T=1$) pairing hamiltonian in LS-coupling,
which is important for heavy N=Z nuclei, is solvable in terms of a $O(8)$
algebra for some special values of the mixing parameter that measures the
competition between $T=0$ and $T=1$ pairing. The $O(8)$ algebra is generated,
amongst others, by the $S=1,T=0$ and $S=0,T=1$ pair creation and annihilation
operators . Shell model algebras,  with only number conserving operators,
that are complementary to the $O(8) \supset O_{ST}(6) \supset O_S(3) \otimes
O_T(3)$, $O(8) \supset [ O_S(5) \supset O_S(3) ] \otimes O_T(3)$ and $O(8)
\supset [ O_T(5) \supset O_T(3)] \otimes O_S(3)$ sub-algebras are
identified.  The problem of classification of states for a given number of
nucleons (called `plethysm' problem in group theory), for these group chains
is solved explicitly for states with $O(8)$ seniority $v=0, 1, 2, 3$ and $4$.
Using them, band structures in isospin space are identified for states with 
$v=0, 1, 2$ and $3$. 

\end{abstract}

\pacs{21.60.Cs, 21.60.Fw, 21.10.Re}

\newpage
\narrowtext

\keywords{N=Z nuclei, Shell model, Spectrum generating algebras,
$O(8)$ algebra, Wigner's $SU(4)$, plethysm, Classification of states}

\section{INTRODUCTION}

In the last few years study of the structure of heavy  odd-odd N=Z nuclei 
(with A $\ge$ 60) near the proton drip line has become an area of intense
research as these nuclei are expected to give new insights into
neutron-proton (np) correlations (or isospin $T=0$ vs $T=1$ pairing) that are
hitherto unknown.  With the development of radioactive ion beam  (RIB)
facilities and  large detector arrays, there are now experimental results for
the energy spectra of $^{62}$Ga, $^{66}$As, $^{70}$Br and $^{74}$Rb
\cite{Ru-96} with many isospin $T=0$ and $T=1$ levels identified; in future 
it is expected that many spectroscopic details of these nuclei will be
available. As a result of this, there are several attempts to develop models
based on shell model and mean-field methods \cite{De-97} for understanding
and predicting the spectroscopic properties of these and other N=Z odd-odd
nuclei in the A=60-100 region. On the other hand models based on group
theory, which gave deeper insights into  the structure  of 'normal' nuclei
\cite{Ta-93}, have also started receiving attention. For example  there are
also attempts to develop the symmetry schemes of the algebraic interacting
boson model with spin($S$)-isospin($T$) degrees of freedom
\cite{IBM-4,Ko-00}  as they will give analytical insights into the structure
of these nuclei. The focus in the present paper is on the symmetry schemes of
the fully fermionic shell model.

With $m$-nucleons in shell model orbits $(j_1,j_2, \ldots)$, the spectrum
generating algebra (SGA) is $U(2\Omega^\prime)$ where 
$\Omega^\prime=\sum_i\;(2j_i+1)$ and 2 comes from isospin. A subalgebra that
conserves isospin  is  $U(2\Omega^\prime) \supset [ U(\Omega^\prime) \supset
Sp(\Omega^\prime)]\otimes SU_T(2)$ and here $Sp(\Omega^\prime)$ corresponds
to pairing. For identical particles, as it is well known, this algebra
corresponds to quasi-spin  $SU(2)$ algebra which generates the seniority
quantum number \cite{Ta-93}. However for  nucleons $Sp(2\Omega^\prime)$
corresponds to $O(5)$ and the $O(5)$ algebra is generated only by the
isovector pair creation and destruction operators, isospin and the number
operators. Hecht and others developed the $O(5)$ algebra in 60's and recently
it is being applied to N=Z nuclei \cite{He-65}.   An unsatisfactory aspect of
the $O(5)$ model is that it does not contain isoscalar pair operator in its
algebra. Therefore it is important to look for an algebra that contains both
$T=0$ and $T=1$ pair operators. To this end it is necessary to consider shell
model in LS-coupling.

Within the nuclear shell model, given the hamiltonian to be a sum of
isoscalar and isovector pairing hamiltonians in LS-coupling for the nucleons,
it is solvable in some special situations using O(8) algebra and its
subalgebras generated amongst others, by the isoscalar ($T=0$) and isovector
($T=1$) pair creation and destruction operators. This was first shown by
Flowers and Szpikowski in 1964 \cite{Fl-64} for the case with nucleons in a
single $\ell$-shell. The $O(8)$ model admits three group-subgroup chains
\cite{Pa-69,Ev-81,Ev-86,Ko-04}: (i) $O(8) \supset O_{ST}(6) \supset O_S(3)
\otimes O_T(3)$; (ii) $O(8) \supset [ O_S(5) \supset O_S(3)] \otimes O_T(3)$;
(iii) $O(8) \supset [ O_T(5) \supset O_T(3)] \otimes O_S(3)$. Section II
gives a brief discussion of these three fermionic $O(8)$ symmetry schemes.
Classification of the many nucleon states in these three symmetry limits is
known only for the $O(8)$ seniority (defined ahead) quantum number $v=0$
\cite{Fl-64,Pa-69,Ev-81,Ev-86,Ko-04}  (for chain (i) results for $v=1,2$ are
also known \cite{Pa-69,He-85,Ko-04}).  Before proceeding further it is
important to mention that, as group theory is formidable, recently the Dyson
boson mapping is being used \cite{Dyson} to gain new insights into the $O(8)$
model as applied to heavy  $N \sim Z$ nuclei. 

In $LS$-coupling, with $\Omega=\sum_i\,(2\ell_i+1)$ number of spatial 
degrees of freedom generated by nucleons in $\ell_1$, $\ell_2$, $\ldots$
orbits, the shell model spectrum generating algebra is $U(4\Omega)$ and all
its generators are number conserving unlike the $O(8)$ algebra. Therefore it
is important to identify the `complementary' (a notion introduced by
Moshinsky and Quesne \cite{Mo-69})  number conserving algebras within
$U(4\Omega)$, with several $\ell$-orbits,  corresponding to the three limits
of the $O(8)$ model. These algebras, identified in Section III and first
reported in \cite{Ko-04}, are: (A) $U(4\Omega) \supset [U(\Omega)  \supset
O(\Omega) \supset O_L(3)] \otimes [O_{ST}(6) \supset O_S(3) \otimes 
O_T(3)]$; (B) $U(4\Omega) \supset [U(2\Omega) \supset Sp(2\Omega) \supset 
\{\l[O(\Omega) \supset O_L(3)] \otimes SU_T(2) \r\}] \otimes SU_S(2)$; (C)
$U(4\Omega) \supset [U(2\Omega) \supset Sp(2\Omega) \supset  \{\l[O(\Omega)
\supset O_L(3)] \otimes SU_S(2) \r\}] \otimes SU_T(2)$. In the reminder of
this paper $O_L(3)$ in these group chains is dropped as the results for
$O(\Omega) \supset O_L(3)$ sector depend explicitly on the $\ell$-orbits
appropriate for a  given nucleus (see however Section VII). 
Note that the chains (A), (B) and (C)
correspond to the chains (i), (ii) and (iii) respectively of the $O(8)$
model. To the extent that the isoscalar plus isovector pairing is important
(i.e. the $O(8)$ limits are fairly good) for heavy $N \sim Z$ nuclei, it is
expected that for nuclei with even number of nucleons $O(8)$ states with $v=
0,2,4$ and for odd mass nuclei with $v=1,3$ will be important. The
purpose of this paper is to give complete classification of states in the
$O(8)$ model, i.e. for the shell model algebras (A), (B) and (C) for 
$v \leq 4$ and thus solving a long standing problem.  These basis states are
associated  with the irreducible representations (irreps) of the groups in
the three  chains. Thus the basic problem is to find the branching rules for
irreps of the various group algebras in each chain into irreps of its
subalgebras. A powerful way of finding such branching rules is provided by 
`plethysms' in group theory. A review of plethysm and its properties can be
found in Refs. \cite{Wy-70,Vladas_book}. They are  applied to some of the
symmetry schemes of the shell model and the interacting boson model of nuclei
in the past \cite{Ko-old,Ko-00}. However recently there is a new systematic
discussion of the plethysm method, development of associated computer codes
and some applications to the interacting boson models of atomic nuclei by one
of the authors \cite{Us1,Us}. In the present paper, following these recent
developments on plethysms \cite{Us1,Us},  we have applied this method
successfully to the chains (A), (B) and (C) and the results are given in
sections IV, V and VI.  Finally Section VII gives conclusions. 

\section{$O(8)$ SYMMETRY SCHEMES}

Let us begin with $m$ nucleons in several $\ell$ orbits $\ell_1$, $\ell_2$,
$\ldots$. Then the single nucleon states are $a^\dagger_{\ell m_\ell\, ;\,
\spin m_s  \,;\,\spin m_t}\,\l|\l. 0 \ran \r.$ and they are $4\Omega$ in
number where  $\Omega=\sum_i\,(2\ell_i+1)$. For a single $\ell$-orbit, pair
states are  defined by two nucleon states with orbital angular momentum zero
($L=0$).  Then by antisymmetry, two nucleon pair states have spin($S$) and
isospin ($T)$ to be $(ST)=(10) \oplus (01)$. With this, the isoscalar and
isovector pair creation operators $D^\dagger_\mu(\ell)$ and
$P^\dagger_\mu(\ell)$  respectively are
\be
D^\dagger_\mu(\ell) = \dis\sqrt{\dis\frac{2\ell +1}{2}}
\l(a^\dagger_{\ell \spin \spin}
a^\dagger_{\ell \spin \spin}\r)^{0,1,0}_{0,\mu,0}\,,\,\;
P^\dagger_\mu(\ell) = \dis\sqrt{\dis\frac{2\ell +1}{2}}
\l(a^\dagger_{\ell \spin \spin}
a^\dagger_{\ell \spin \spin}\r)^{0,0,1}_{0,0,\mu}\;. 
\ee
Note that we are using $(L,S,T)$ order in (1). For the multi-orbit case 
one can define the generalized isoscalar and isovector pair operators
$D^\dagger_\mu$ and $P^\dagger_\mu$ as linear combinations of single orbit
$D^\dagger_\mu(\ell)$'s and $P^\dagger_\mu(\ell)$'s respectively except for
phase factors,
\be
D^\dagger_\mu = \dis\sum_{\ell}\, \beta_\ell \;D^\dagger_\mu(\ell)\;,
\;\;\;
P^\dagger_\mu = \dis\sum_{\ell}\, \beta_\ell \;P^\dagger_\mu(\ell)\;;
\;\;\;\beta_\ell=+1\;\;\mbox{or}\;\;-1\;. 
\ee
Now it is possible to define the pairing hamiltonian in $LS$-coupling,
\be
H_{pairing}(x) = -(1-x)\, \dis\sum_{\mu}\,P^\dagger_\mu P_\mu\; 
- \; (1+x) \, \dis\sum_{\mu}\,D^\dagger_\mu D_\mu\;. 
\ee
Note that $P_\mu = (P^\dagger_\mu)^\dagger$ and 
$$
P_\mu(\ell) = \l(P^\dagger_\mu(\ell)
\r)^\dagger = (-1)^\mu \sqrt{(2\ell +1)/2}\; ({\tilde{a}}_{\ell \spin \spin}
{\tilde{a}}_{\ell \spin \spin})^{0,0,1}_{0,0, -\mu}
$$
where $\tilde{a}$ is related to $a$ by 
$$
a_{\ell m_\ell\, ;\, \spin m_s  \,;\,\spin
m_t}=(-1)^{\ell+1+m_\ell -m_s -m_t}{\tilde{a}}_{\ell -m_\ell\, ; \, \spin
-m_s \,;\,\spin -m_t}\;.
$$
Similarly $D_\mu$ and $D_\mu(\ell)$ are defined. At
this stage it is also necessary to define spin ($S^1_\mu$), isospin
($T^1_\mu$), Gamow-Teller ($(\sigma \tau)^{1,1}_{\mu,\mu^\prime}$) and 
number ($\hat{n}$ or the equivalent $Q_0$) operators,
\be
\barr{l}
X^{S,T}_{\mu,\mu^\prime} = \dis\sum_{\ell} \dis\sqrt{2\ell + 1} 
\l(a^\dagger_{\ell \spin 
\spin} {\tilde{a}}_{\ell \spin \spin}\r)^{0,S,T}_{0,\mu,\mu^\prime}\;, \nn8
S^1_\mu=X^{1,0}_{\mu,0}\,,\;\;
T^1_\mu = X^{0,1}_{0,\mu}\,,\;\;(\sigma \tau)^{1,1}_{\mu,\mu^\prime} =
X^{1,1}_{\mu,\mu^\prime}\;, \nn8
\hat{n} = 2\,X^{0,0}_{0,0}\,,\;\;\; Q_0=\frac{\hat{n}}{2}-\Omega\;.
\earr
\ee
By evaluating the commutators, it can be shown that the 28 operators 
$P^\dagger_\mu$, $P_\mu$, $D^\dagger_\mu$, $D_\mu$,
$S^1_\mu$, $T^1_\mu$, $(\sigma \tau)^{1,1}_{\mu,\mu^\prime}$ and $Q_0$  
generate the following algebras:
\be
\barr{l}
O(8)\;\;:\;\;\{ P^\dagger_\mu,\; P_\mu,\; D^\dagger_\mu,\; D_\mu,\;
S^1_\mu,\; T^1_\mu,\; (\sigma \tau)^{1,1}_{\mu,\mu^\prime},\; Q_0\} \\
O_{ST}(6) \;\;:\;\;\{S^1_\mu,\; T^1_\mu,\; (\sigma \tau)^{1,1}_{
\mu,\mu^\prime}\} \\
O_S(5) \;\;: \;\; \{D^\dagger_\mu,\; D_\mu,\; S^1_\mu,\; Q_0\} \\ 
O_T(5) \;\;: \;\;\{P^\dagger_\mu,\; P_\mu,\; T^1_\mu,\; Q_0\}\\
O_S(3) \;\;:\;\;S^1_\mu \\
O_T(3)\;\; :\;\; T^1_\mu\;.
\earr
\ee
The $O_{ST}(6)$ algebra in (5) is nothing but Wigner's spin-isospin
supermultiplet $SU_{ST}(4)$ algebra \cite{Wi-37}. Note that $O_S(3) \sim
SU_S(2)$, $O_T(3) \sim SU_T(2)$ and $SU_{ST}(4) \supset SU_S(2)  \otimes
SU_T(2)$. Quadratic Casimir  operators of the groups in (5) are (besides
$S^2$ for $O_S(3)$ and $T^2$ for  $O_T(3)$),
\be
\barr{l}
C_2(O(8)) \;\;=\;\; 2\l(\dis\sum_\mu P^\dagger_\mu P_\mu +  
\dis\sum_\mu D^\dagger_\mu D_\mu\r) + C_2(O(6)) + Q_0(Q_0-6)\;, \\
C_2(O_{ST}(6)) \;\;= \;\; S^2 + T^2 + (\sigma \tau) \cdot (\sigma \tau)\;, \\
C_2(O_S(5)) \;\;= \;\;2 \dis\sum_\mu D^\dagger_\mu D_\mu + S^2 +
Q_0(Q_0-3)\;,\\
C_2(O_T(5))\;\; =\;\; 2 \dis\sum_\mu P^\dagger_\mu P_\mu + T^2 + Q_0(Q_0-3)
\;.
\earr
\ee
In (6), the dot-product is defined by $A^k \cdot B^k = (-1)^k \sqrt{2k+1} 
(A^k B^k)^0$ and similarly $A^{k_1 k_2} \cdot B^{k_1 k_2}$ is defined.
It is seen from (5) that the pairing hamiltonian (3), for any $x$,
has $O(8)$ symmetry as it contains only the generators of $O(8)$. Moreover,
by examining the quadratic Casimir invariants in (6), it is seen that the
pairing hamiltonian (3) is solvable in the situations $x=0,1,-1$ and the
corresponding subalgebras (group-subgroup chains) of $O(8)$ are,
\be
\barr{rcl} 
x=0 & : & O(8) \supset O_{ST}(6) \supset O_S(3) \otimes O_T(3) \\
x=1 & : & O(8) \supset \l[O_S(5) \supset O_S(3)\r] \otimes O_T(3) \\
x=-1 & : & O(8) \supset \l[O_T(5) \supset O_T(3)\r] \otimes O_S(3) 
\earr
\ee
All these group-subgroup chains have number non-conserving operators.

\section{Shell model algebras complementary to the $O(8)$ algebras}
 
We will now consider the $O(8)$ chains in (7) in terms of their
"complementary" number conserving shell model group chains. 
In the $(\ell_1, \ell_2, \ldots)^m$ space, there is a $U(4\Omega)$ algebra
generated by the operators
\be
u^{L,S,T}_{\mu_\ell,\mu_S,\mu_T}(\ell_1, \ell_2) = \l(a^\dagger_{\ell_1 \spin 
\spin} {\tilde{a}}_{\ell_2 \spin \spin}\r)^{L,S,T}_{\mu_\ell,\mu_S,\mu_T}\;.
\ee
With respect to this $U(4\Omega)$, all the $m$-nucleon states behave as the
totally antisymmetric irrep $\{1^m\}$ and this is our starting point.

\subsection{ $U(4\Omega) \supset [U(\Omega) \supset O(\Omega)] \otimes 
[SU_{ST}(4) \supset SU_S(2) \otimes SU_T(2)]$ chain}

With good $(LST)$, $U(4\Omega)$ algebra can be decomposed into a product of
space $U(\Omega)$ and spin-isospin $SU_{ST}(4)$ algebras. This gives the 
group chain,
\be
U(4\Omega) \supset \l[U(\Omega) \supset O(\Omega) \r] \otimes
\l[SU_{ST}(4) \supset SU_S(2) \otimes SU_T(2)\r]\;.
\ee
Note that $SU_{ST}(4) \sim O_{ST}(6)$, $SU_{S}(2) \sim O_{S}(3)$ and
$SU_{T}(2) \sim O_{T}(3)$. Following the results in Appendix A  and B of
\cite{Ko-00} it is straight forward to write down the  generators and the
quadratic Casimir operators ($C_2$'s) of the algebras in (9),
\be
\barr{rcl}
U(4\Omega) & : & u^{L,S,T}_{\mu_\ell,\mu_S,\mu_T}(\ell_1, \ell_2) \nn8
U(\Omega) & : & 2\; u^{L,0,0}_{\mu_\ell,0,0}(\ell_1, \ell_2) \nn8
U_{ST}(4) & : & X^{S,T}_{\mu_S,\mu_T} = \dis\sum_{\ell} \dis\sqrt{2\ell +1}\; 
u^{0,S,T}_{0,\mu_S,\mu_T}(\ell, \ell) \nn8
O_{ST}(6) \sim SU_{ST}(4) & : & X^{S,T}_{\mu_S,\mu_T}\;;\;\;\;
(ST)=(10),(01),(11) \nn8
O(\Omega) & : & 2\; u^{L=odd,0,0}_{\mu_\ell,0,0}(\ell, \ell),\;\;V^L_\mu(
\ell_1, \ell_2)\;\;\mbox{with}\;\;\ell_1>\ell_2\;; \nn8
& & {\barr{l} V^L_\mu(\ell_1,\ell_2)=2\l[\alpha(\ell_1, \ell_2) 
(-1)^{\ell_1+\ell_2+L} \r]^{\spin} \times \\
\l\{u^{L,0,0}_{\mu,0,0}(\ell_1, \ell_2) + \alpha(\ell_1, \ell_2) (-1)^{L}
u^{L,0,0}_{\mu,0,0}(\ell_2, \ell_1)\r\} \earr}\;, \nn8
& & \l|\alpha(\ell_1, \ell_2)\r|^2=1,\;\;\alpha(\ell_1, \ell_2)
\alpha(\ell_2, \ell_3)=-\alpha(\ell_1, \ell_3) \nn8
C_2(U(\Omega)) & = & 4 \dis\sum_{\ell_1, \ell_2, L} (-1)^{\ell_1+\ell_2}
u^{L,0,0}(\ell_1, \ell_2) \cdot u^{L,0,0}(\ell_2, \ell_1)\;, \nn8
C_2(O(\Omega)) & = & 8 \dis\sum_{\ell, L=odd} u^{L,0,0}(\ell, \ell) \cdot 
u^{L,0,0}(\ell, \ell)  \nn8
& & \;\;\;\;\;\;\;\;\;\;\;\;\;\;\;+ 
\dis\sum_{\ell_1 > \ell_2; L} V^L(\ell_1,\ell_2)
\cdot V^L(\ell_1,\ell_2)\;, \nn8
C_2(U_{ST}(4)) & = & \dis\sum_{S,T}\;X^{S,T} \cdot X^{S,T}\;. 
\earr
\ee
It should be noted that $O(\Omega)$ is not unique and in the
multi-orbit case there are several $O(\Omega)$'s as defined by distinct 
$\alpha(\ell_1, \ell_2)$'s in (10). Using (10), it can be proved that,
\be
\barr{rcl}
C_2(U(\Omega)) + C_2(U_{ST}(4)) & = & \hat{n} (4+\Omega)\;, \nn8
C_2(U_{ST}(4)) & = & C_2(O_{ST}(6)) +{\hat{n}}^2/4\;, \nn8
C_2(U(\Omega)) - C_2(O(\Omega)) & = & 2\l[\dis\sum_\mu P^\dagger_\mu P_\mu +  
\dis\sum_\mu D^\dagger_\mu D_\mu\r] + \hat{n}\;.
\earr
\ee
The third equality in (11) is valid only when
\be
\beta_{\ell_1} \beta_{\ell_2} = -\alpha(\ell_1, \ell_2)\;,\;\;\;
\ell_1 \neq \ell_2\;. 
\ee
The relations in (12) with $\beta$'s defining the pair operators
in multi-orbit case (see Eq. (2)) and $\alpha$'s defining $O(\Omega)$
generators (see Eq. (10)), via $C_2(O(8))$ given in (6),  
connect $O(8)$ with $O(\Omega)$. In fact using (6,11) it is seen that,
\be
C_2(O(8))=-C_2(O(\Omega)) + \Omega(\Omega+6)\;,
\ee
\be
\dis\sum_\mu P^\dagger_\mu P_\mu +  \dis\sum_\mu D^\dagger_\mu D_\mu = 
\spin \l\{-C_2(O(\Omega)) -C_2(O_{ST}(6)) -Q_0(Q_0-6) +
\Omega(\Omega+6)\r\}\;.
\ee
Thus, clearly the chain (9) is equivalent to the $O(8) \supset O_{ST}(6) 
\supset O_S(3) \otimes O_T(3)$ chain and it solves the pairing hamiltonian 
(3) for $x=0$. One important result that follows from  (2,10,12) is that in
the multi-orbit case, there are multiple definitions of pair operators $P$
and $D$ as given by (2) and for each of these definitions there is a unique
$O(\Omega)$ as defined by (10,12). In all the previous studies involving 
several orbits, the choice $\beta_\ell=1$ is made.  Also $O(8)$ will not
allow for solving the isovector plus isoscalar pairing hamiltonian with
$\beta$'s different for the isoscalar and isovector parts. 

In order to construct the spectra generated by the group chain (9), we will
now turn to the irreps of the groups in (9) and their reductions. Throughout
this article, we use Wybourne's \cite{Wy-70} notations $\{--\}$, $[---]$ and
$\lan --- \ran$ respectively for denoting $U(N)$, $O(N)$ and $Sp(N)$ irreps.

Our starting point is $\{1^m\}$ irrep of $U(4\Omega)$. Its reduction to
irreps of $U(\Omega)$ in (9) is simple as $U(\Omega)$ appears as a factor in
the direct product subgroup with other group being $SU_{ST}(4) \sim
O_{ST}(6)$ (or $U_{ST}(4)$). The $U_{ST}(4)$ irreps $\{f\}=\{f_1 f_2 f_3 
f_4\}$ uniquely define (by transposition) the $U(\Omega)$ irreps
\cite{Wy-70},
\be
\barr{rcl}
U(\Omega) &:& \{\tilde{f}\} = \l\{4^{f_4} 3^{f_3-f_4} 2^{f_2-f_3} 1^{f_1-f_2
}\r\}\;; \nn8
& & \dis\sum_i f_i = m,\;\;f_1 \geq f_2 \geq f_3 \geq f_4 \geq 0 \nn8
O_{ST}(6) &:& \l[P_1, P_2, P_3\r] = \nn8
& & \l[\dis\frac{f_1+f_2-f_3-f_4}{2}, 
\dis\frac{f_1-f_2+f_3-f_4}{2}, \dis\frac{f_1-f_2-f_3+f_4}{2} \r]\;.
\earr
\ee
Given in (15) are also the equivalent $O_{ST}(6)$ irreps $[P_1, P_2, P_3]$ 
in terms of the $U_{ST}(4)$ irrep labels. 
Analogous to the $U(\Omega)$ irreps, one can write the $O(\Omega)$ irreps by
introducing the quantum numbers $v$ and $[p_1 p_2 p_3]$ as (see \cite{Fl-64}),
\be
\barr{rcl}
O(\Omega) &:& [\tilde{\mu}] = [4^{\mu_4} 3^{\mu_3-\mu_4} 2^{\mu_2-\mu_3}
1^{\mu_1-\mu_2}] \Leftrightarrow v,\l[p_1, p_2, p_3\r]\;, \nn8
& & v=\dis\sum_i \mu_i\,,\;\;\;\mu_1 \geq \mu_2 \geq \mu_3 \geq \mu_4 \geq
0,\nn8 
& & \l[p_1, p_2, p_3\r] = \nn8
& & \l[\dis\frac{\mu_1+\mu_2-\mu_3-\mu_4}{2}, 
\dis\frac{\mu_1-\mu_2+\mu_3-\mu_4}{2}, \dis\frac{\mu_1-\mu_2-\mu_3+
\mu_4}{2} \r]\;.
\earr
\ee
With $\{1^m\}_{U(4\Omega)} \rightarrow \{\tilde{f}\}_{U(\Omega)} \otimes
\l[P_1P_2P_3\r]_{O_{ST}(6)}$, the important reduction that is needed is 
$\{\tilde{f}\}_{U(\Omega)} \rightarrow v\l[p_1p_2p_3\r]$ of $O(\Omega)$
(of-course $\l[P_1P_2P_3\r] \rightarrow (ST)$ and $v,\l[p_1p_2p_3\r] 
\rightarrow L$ are also needed). Before addressing this problem in Section IV, 
let us examine the matrix elements of the quadratic Casimir invariants. Using 
the general results
\be
\barr{rcl}
\lan C_2(U(N))\ran^{\{F_1 F_2, \ldots\}} & = & \dis\sum_i F_i(F_i+
N+1-2i)\;, \nn8
\lan C_2(O(N))\ran^{[\omega_1 \omega_2, \ldots]} & = & \dis\sum_i 
\omega_i(\omega_i+N-2i)
\earr
\ee
it is seen that
\be
\barr{rcl}
\lan C_2(O(6))\ran^{\l[P_1P_2P_3\r]} & = & P_1(P_1+4) + P_2(P_2+2) + 
P_3^2\;, \nn8
\lan C_2(O(\Omega))\ran^{v,\l[p_1 p_2 p_3\r]} & = & v(\Omega+3-v/4) \nn8
& & - \l[p_1(p_1+4) + p_2(p_2+2) + p_3^2\r] \nn8
\Rightarrow \lan C_2(O(8))\ran^{v,\l[p_1 p_2 p_3\r]} & = & Q(Q+6) +
\l[p_1(p_1+4) + p_2(p_2+2) + p_3^2\r]\;.
\earr
\ee
where $Q=\Omega-v/2$. The last equality follows from (13).  From (14,18) it is
clear that the states with $v=0$ will be lowest in energy for the pairing
hamiltonian (3) with $x=0$. Therefore it is meaningful to consider $v \leq 4$
states and workout the allowed $O_{ST}(6)$ irreps. This exercise is carried out
in Section IV.  Using Eqs. (14) and (18) the spectrum can be constructed. The
energies are  given by
\be
\barr{l}
\lan H_{pairing}(x=0)\ran^{m,v,[p],[P],(ST)} = 
\spin \l\{-\frac{1}{4}\l(m-v\r)\l(4\Omega+12 -m -v \r) \r.\nn8
- \l. \l[p_1(p_1+4) + p_2(p_2+2) + p_3^2\r] + \l[P_1(P_1+4) + P_2(P_2+2) 
+ P_3^2\r] \r\}
\earr
\ee
and it is independent of the $(ST)$ quantum numbers. 

\subsection{$U(4\Omega) \supset \l[ U(2\Omega)\supset Sp(2\Omega) \supset 
O(\Omega) \otimes SU_T(2)\r] \otimes SU_S(2)$ chain}

In order to identify the number conserving group chain that is complementary
to $O(8) \supset \l[O_S(5) \supset O_S(3)\r] \otimes O_T(3)$, obviously one
has to start with the $U(2\Omega) \otimes SU_S(2)$ subalgebra of $U(4\Omega)$
algebra. As $U(2\Omega)$ contains $O(\Omega) \otimes SU_T(2)$ as a subalgebra,
for completing the group chain, one has to find the subalgebras between these 
two algebras. It is easy to see that $Sp(2\Omega)$
is the subalgebra one is looking for. Therefore the complementary
group-subgroup chain is,
\be
U(4\Omega) \supset \l[U(2\Omega) \supset Sp(2\Omega) \supset \l\{O(\Omega)
\otimes SU_T(2) \r\}\r] \otimes SU_S(2)\;.
\ee  
To establish this result let us consider the generators and the quadratic 
Casimir operators ($C_2$'s) of the algebras in (20),
\be
\barr{rcl}
U(2\Omega) & : & \dis\sqrt{2}\; u^{L,0,T}_{\mu_\ell,0,\mu_T}(\ell_1, 
\ell_2) \nn8
U_S(2) & : & Y^S_{\mu_S} = \dis\sum_{\ell} \dis\sqrt{2(2\ell +1)}\; 
u^{0,S,0}_{0, \mu_S,0}(\ell, \ell) \nn8
Sp(2\Omega) & : & \dis\sqrt{2}\; u^{L,0,T}_{\mu_\ell,0,\mu_T}(\ell, 
\ell)\,;\;L+T=odd,\;V^{L,T}_{\mu , \mu_T}(\ell_1,
\ell_2)\;\;\mbox{with}\;\;\ell_1>\ell_2, \nn8
& & {\barr{l} V^{L,T}_{\mu, \mu_T}(\ell_1,\ell_2)=\dis\sqrt{2}
\l[\alpha(\ell_1, \ell_2) (-1)^{\ell_1+\ell_2+L+T} \r]^{\spin} \times \\
\l\{u^{L,0,T}_{\mu,0,\mu_T}(\ell_1, \ell_2) + 
\alpha(\ell_1, \ell_2) (-1)^{L+T}
u^{L,0,T}_{\mu,0,\mu_T}(\ell_2, \ell_1)\r\} \earr}\;, \nn8
& & \l|\alpha(\ell_1, \ell_2)\r|^2=1,\;\;\alpha(\ell_1, \ell_2)
\alpha(\ell_2, \ell_3)=-\alpha(\ell_1, \ell_3) \nn8
C_2(U(2\Omega)) & = & 2 \dis\sum_{\ell_1, \ell_2, L, T} (-1)^{\ell_1+\ell_2}\;
u^{L,0,T}(\ell_1, \ell_2) \cdot u^{L,0,T}(\ell_2, \ell_1)\;, \nn8
C_2(Sp(2\Omega)) & = & 4 \dis\sum_{\ell, L+T=odd} u^{L,0,T}(\ell, \ell) 
\cdot u^{L,0,T}(\ell, \ell) \nn8
& & \;\;\;\;\;\;\;\;\;\;\;\;+ \dis\sum_{\ell_1 > \ell_2; L,T} 
V^{L,T}(\ell_1,\ell_2) \cdot V^{L,T}(\ell_1,\ell_2)\;, \nn8
C_2(U_S(2)) & = & \dis\sum_{S}\;Y^S \cdot Y^S \;,
\earr
\ee
and the $O(\Omega)$ generators and quadratic Casimir operator are given by 
Eq.(10). It should be noted that again $Sp(2\Omega)$ is not unique in the
multi-orbit case and just as $O(\Omega)$, it is defined by distinct 
$\alpha(\ell_1, \ell_2)$'s in (21). Using (21), it can be proved that,
\be
\barr{rcl}
C_2(U(2\Omega)) + C_2(U_S(2)) & = & \hat{n} (2+2\Omega)\;, \nn8
C_2(U_S(2)) & = & 2S^2 +{\hat{n}}^2/2\;, \nn8
C_2(U(2\Omega)) - C_2(Sp(2\Omega)) & = & 4\l[\dis\sum_\mu D^\dagger_\mu 
D_\mu\r] - \hat{n}\;.
\earr
\ee
The third equality in (22) is valid only when (12) is satisfied. Comparing
(22) with (6), it is seen that $Sp(2\Omega)$ in (20) is related to $O_S(5)$, 
\be
C_2(O_S(5))=-\spin C_2(Sp(2\Omega)) + \Omega(\Omega+3)\;,
\ee
\be
2 \dis\sum_\mu D^\dagger_\mu D_\mu = -\spin 
C_2(Sp(2\Omega)) -S^2 -Q_0(Q_0-3) + \Omega(\Omega+3)\;.
\ee
Thus, clearly the chain (20) is equivalent to the $O(8) \supset \l[O_S(5) 
\supset O_S(3)\r] \otimes O_T(3)$ chain and it solves the pairing 
hamiltonian (3) for $x=1$.

In order to construct the spectra generated by the group chain (20), we will
now turn to the irreps of the groups in (20). Just as
before, the starting point is $\{1^m\}$ irrep of $U(4\Omega)$. 
Its reduction to $U(2\Omega)$  irreps is simple as $U(2\Omega)$ in (20) 
appears in the direct product subgroup with  the other group being 
$SU_S(2)$ (or $U_S(2)$). The $U_S(2)$ irreps $\{f_S\}=\{f_1 f_2 \}$ uniquely
define (by transposition) the $U(2\Omega)$ irreps,
\be
U(2\Omega) \;:\;\;\l\{\tilde{f}_S\r\}=\l\{2^{f_2} 1^{f_1-f_2}\r\}\;;\;
f_1+f_2=m,\;\;f_1 \ge f_2 \geq 0,\;S=(f_1-f_2)/2\;.
\ee
Now the $Sp(2\Omega)$ irreps can be written as,
\be
Sp(2\Omega) \;:\;\;\lan\tilde{\mu}_S\ran=\lan 2^{\mu_1} 1^{\mu_2}\ran\;;
\;\;\;v_S=2\mu_1+\mu_2\;,\;\;t_S=\dis\frac{\mu_2}{2}\;.
\ee
The $v_S$ and $t_S$ quantum numbers are introduced by
examining the eigenvalue expression for $C_2(Sp(2\Omega))$,
\be
\barr{l}
\lan C_2(Sp(2\Omega)) \ran^{\lan \lambda_1 \lambda_2 \ldots \ran} =
\dis\sum_i\;\lambda_i\l(\lambda_i + 2\Omega +2 -2i\r) \nn8
\Rightarrow \;\; \lan C_2(Sp(2\Omega)) \ran^{\lan2^{\mu_1} 1^{\mu_2}\ran} =
\nn8
2\l[\Omega(\Omega+3) - \l(\Omega -\dis\frac{v_S}{2}\r) \l(\Omega -
\dis\frac{v_S}{2}+3\r) -t_S(t_S+1)\r]\;.
\earr
\ee
Using Eqs. (24) and (27) we have,
\be
\barr{rcl}
\lan H_{pairing}(x=1)\ran^{m,S,v_S,t_S,T,v,[p]} & = & -\frac{1}{4} 
\l(m-v_S\r) \l(4\Omega+6-m-v_S\r) \nn8
& & - t_S(t_S+1) + S(S+1)
\earr
\ee
and thus the energies in the limit (20) do not depend (explicitly)
on $T$, $v$ and $[p]$ quantum numbers. Eq. (28) shows that
for large $\Omega$ and fixed $S$, smallest $v_S$ states will be lowest in 
energy for the chain (20). 

\subsection{$U(4\Omega) \supset \l[U(2\Omega)\supset Sp(2\Omega) 
\supset O(\Omega) \otimes SU_S(2) \r] \otimes SU_T(2)$ chain}

The chain complementary to $O(8) \supset \l[O_T(5) \supset O_T(3)\r] \otimes
O_S(3)$ is 
\be
U(4\Omega) \supset \l[U(2\Omega)\supset Sp(2\Omega) 
\supset O(\Omega) \otimes SU_S(2) \r] \otimes SU_T(2)\;. 
\ee
All results for this chain follow from Section III.B by simply interchanging 
$S \Leftrightarrow T$. Thus the $U(2\Omega)$ and $Sp(2\Omega)$ algebras in Eq.
(29) are defined in orbital-spin space and the $Sp(2\Omega)$ irreps are
labeled by $(v_T,t_T)$. The energy formula here is,
\be
\barr{rcl}
\lan H_{pairing}(x=-1)\ran^{m,S,v_T,t_T,T,v,[p]} & = & -\frac{1}{4} 
\l(m-v_T\r) \l(4\Omega+6-m-v_T\r) \nn8
& & - t_T(t_T+1) + T(T+1)\;.
\earr
\ee
Therefore chain (29) generates, with the $T(T+1)$ term in Eq. (30), rotational
spectra in isospace and thus it is different from chain (20). 

\section{Irrep reductions for the chain $U(4\Omega) \supset [U(\Omega) 
\supset O(\Omega)] \otimes [SU_{ST}(4) \supset SU_S(2) \otimes SU_T(2)]$}

For the $U(4\Omega) \supset [U(\Omega) \supset O(\Omega)] \otimes SU_{ST}(4)$
chain (hereafter called Limit-I), branching rules for $U(\Omega) 
\supset O(\Omega)$ and $SU_{ST}(4) \supset
SU_S(2) \otimes SU_T(2)$ (or equivalently $O_{ST}(6) \supset O_S(3) \otimes
O_T(3)$) are given in subsections IV.A and IV.B respectively. As discussed in
Section III.A, our interest is in obtaining the irreps that belong to $O(8)$ 
seniority $v \leq 4$.

\subsection{$U(\Omega)$ and $O_{ST}(6)$ irreps for $O(8)$ seniority 
$v \leq 4$}

The branching rules for $U(\Omega)  \supset O(\Omega)$ are obtained by finding
which 4-column $U(\Omega)$ irreps $\{\tilde{f}\}$ of degree $m$ have a given
$O(\Omega)$ irrep  $[\tilde{\mu}]$ of degree $v$ in their restriction to
$O(\Omega)$ . They follow from the rule \cite{Wy-70},
\be
\barr{rcl}
\{\tilde{f}\} & = & \dis\sum_{\{\tilde{\mu}\}} 
\dis\sum_{\{\tilde{\delta}\}_{even}} \; 
\Gamma_{\{\tilde{\delta}\}\{\tilde{\mu}\}\{\tilde{f}\}}[\tilde{\mu}]\;, \nn8
\{\tilde{\delta}\} & = & \{0\}, \{2\}, \{4\}, \{2^2\}, \{6\}, \{42\}, 
\{2^3\}, \ldots
\earr  
\ee
In (31) $\Gamma_{\{\tilde{\delta} \}\{\tilde{\mu}\}\{\tilde{f}\}}$ is the
multiplicity of $\{\tilde{f}\}$  in the reduction
$\{\tilde{\delta}\}\,\{\tilde{\mu}\} \rightarrow  \{\tilde{f}\}$. Therefore,
the $\{\tilde{f}\}$ irreps we are looking for are those produced  by the
outer product of the given irrep $\{\tilde{\mu}\}$ by all allowed  irreps
$\{\tilde{\delta}\}$ with even entries.  The irrep $\{\tilde{f}\}$ being a
4-column irrep, so must be $\{\tilde{\delta} \}$ and $\{\tilde{\mu}\}$, then
$\{\tilde{\delta}\}=\{4^r,2^s\}$ with  $4r+2s+v=m$. Note that $v$ is the
$O(8)$ seniority quantum number given in Eq. (16). For $v=$ even, one has
$m=$even$=4k$ and $4k+2$ where $k$ is a positive integer (similarly $\ell$
ahead). Then,
\be
\barr{l}
m=4k \Rightarrow s=2(k-r)-v/2 \Rightarrow    \\
s=0,2,4,...,2k-v/2 \quad \mbox{ for } v=4\ell    \\
s=1,3,5,...,2k-v/2 \quad \mbox{ for } v=4\ell+2  \nn8
m=4k+2 \Rightarrow s=2(k-r)+1-v/2 \Rightarrow \\
s=1,3,5,...,2k+1-v/2 \quad \mbox{ for } v=4\ell \\
s=0,2,4,...,2k+1-v/2 \quad \mbox{ for } v=4\ell+2\;.  
\earr
\ee
For $v=$ odd, one has $m=$odd$=4k+1$ and $4k+3$. This gives,
\be
\barr{l}
m=4k+1 \Rightarrow s=2(k-r)-(v-1)/2 \Rightarrow \\
s=0,2,4,...,2k-(v-1)/2 \quad \mbox{ for } v=4\ell+1  \\
s=1,3,5,...,2k-(v-1)/2 \quad \mbox{ for } v=4\ell+3   \nn8
m=4k+3 \Rightarrow s=2(k-r)-(v-3/2) \Rightarrow  \\
s=1,3,5,...,2k-(v-3)/2 \quad \mbox{ for } v=4\ell+1    \\
s=0,2,4,...,2k-(v-3)/2 \quad \mbox{ for } v=4\ell+3\; . 
\earr
\ee
Note that $s=0 \Rightarrow [P]=[p]$. Using the rules of outer products, the 
results for $v=0,1,2,3$ and 4 are obtained and they are as follows:

\noindent {\Large{$(i)\;$\bf{$v = 0$}}}

\noindent From  Eqs.(31) and (32) one has easily,
\be
\barr{l}
[\tilde{\mu}]=[0] \Rightarrow [\mu]=[0] \Rightarrow 
\l[p\r]=\l[0\r] \Rightarrow \l\{\tilde{f}\r\}=\l\{4^r,2^s\r\} \\ 
\Rightarrow  \{f\}=\{r+s,r+s,r,r\} \Rightarrow 
\l[P\r]=\l[s,0,0\r]\;; \\    
s=0,2,4,...,2k \quad \mbox{ for } m=4k \\    
s=1,3,5,...,2k+1 \;\;\mbox{ for } m=4k+2\; .  
\earr
\ee

\noindent {\Large{$(ii)\;$\bf{$v=1$}}}

\noindent From Eq. (33) one has,
$$
\barr{l}
s=0,2,4,...,2k   \quad \mbox{ for } m=4k+1 \\
s=1,3,5,...,2k+1 \quad \mbox{ for } m=4k+3\;.  
\earr
$$

\noindent $(ii.1): \;[\tilde{\mu}]=[1] \Rightarrow [\mu]=[1,0,0,0] 
\Rightarrow [p] = [1/2,1/2,1/2]$. 

\be
\barr{l}
\{\tilde{f}\}=\l\{4^r,3,2^{s-1}\r\}_{s \geq 1}+\l\{4^r,2^s,1\r\}_{s \geq 0} 
\Rightarrow  \\
\{f\}=\{r+s,r+s,r+1,r\}+\{r+s+1,r+s,r,r\}   \Rightarrow  \\
\l[P\r]=\l[s-1/2,1/2,-1/2\r]+\l[s+1/2,1/2,1/2\r] \;.
\earr
\ee
In Eq. (35) the notation $\{\tilde{f}\}_{s \geq s_0}$ means that the irrep 
$\{\tilde{f}\}$ is present only when $s \geq s_0$ and the same applies to
$\{f\}$ and $[P]$ produced by it. This notation is used in the remaining 
part of this section.

\noindent{\Large{$(iii)\;$\bf{$v= 2$}}}

\noindent From Eq. (32) one has, 
$$ 
\barr{l} s=1,3,...,2k-1, \quad \mbox{ for }   m=4k \\
s=0,2,4,...,2k  \quad \mbox{ for }   m=4k+2 \;.  
\earr 
$$ 

\noindent $(iii.1):\;[\tilde{\mu}]=[2] \Rightarrow [\mu] =
[1,1,0,0] \Rightarrow \l[p\r]=\l[1,0,0\r]\;.$ 

\be 
\barr{l} 
\l\{\tilde{f}\r\}=\l\{4^{r+1},2^{s-1}\r\}_{s \geq 1} +
\l\{4^r,3,2^{s-1},1\r\}_{s \geq 1} +
\l\{4^r,2^{s+1}\r\}_{s \geq 0}  \Rightarrow \\
\{f\}=\{r+s,r+s,r+1,r+1\}+\{r+s+1,r+s,r+1,r\} \\
+\{r+s+1,r+s+1,r,r\} \\
\Rightarrow [P]=[s-1,0,0]+[s,1,0]+[s+1,0,0]\;.  
\earr 
\ee 

\noindent $(iii.2):\;[\tilde{\mu}]=[1^2] \Rightarrow [\mu]
= [2,0,0,0] \Rightarrow  [p]=[1,1,1]\;.$ 
 
\be 
\barr{l}
\l\{\tilde{f}\r\}=\l\{4^r,3^2,2^{s-2}\r\}_{s \geq 2} +
\l\{4^r,3,2^{s-1},1\r\}_{s \geq 1}
+\l\{4^r,2^s,1^2\r\}_{s \geq 0}  \Rightarrow \\
\{f\}=\{r+s,r+s,r+2,r\}+\{r+s+1,r+s,r+1,r\}+\{r+s+2,r+s,r,r\} \\ 
\Rightarrow [P] = [s-1,1,-1]+[s,1,0]+[s+1,1,1]\;.  
\earr 
\ee

\noindent {\Large{$(iv)\;$\bf{$v = 3$}}}

\noindent From Eq. (33) one has,
$$
\barr{l}
s=1,3,5,...,2k-1, \quad \mbox{ for } m=4k+1 \\
s=0,2,4,...,2k    \quad \mbox{ for } m=4k+3 \;.
\earr
$$

\noindent $(iv.1):\; [\tilde{\mu}]=[3] \Rightarrow [\mu]
= [1,1,1,0] \Rightarrow [p]=[1/2,1/2,-1/2]\;. $

\be
\barr{l}
\{\tilde{f}\}=\l\{4^{r+1},2^{s-1},1\r\}_{s \geq 1}
+\l\{4^r,3,2^s\r\}_{s \geq 0} \Rightarrow \\
\{f\}=\{r+s+1,r+s,r+1,r+1\}+\{r+s+1,r+s+1,r+1,r\} \Rightarrow \\
\l[P\r]=\l[s-1/2,1/2,1/2\r]+\l[s+1/2,1/2,-1/2\r]\;.  
\earr
\ee

\noindent $(iv.2):\; [\tilde{\mu}]=[2,1] \Rightarrow [\mu] =
[2,1,0,0] \Rightarrow [p]=[3/2,1/2,1/2]\;. $

\be
\barr{l}
\{\tilde{f}\}=\l\{4^{r+1},3,2^{s-2}\r\}_{s \geq 2} + 
\l\{4^r,3^2,2^{s-2},1\r\}_{s \geq 2} + \l\{4^{r+1},2^{s-1},1\r\}_{s \geq 1} \\
+\l\{4^r,3,2^s\}_{s \geq 1} +\{4^r,3,2^{s-1},1^2\r\}_{s \geq 1} 
+ \l\{4^r,2^{s+1},1\r\}_{s \geq 0} \Rightarrow \\
\{f\}=\{r+s,r+s,r+2,r+1\}+ \{r+s+1,r+s,r+2,r\} \\ 
+ \{r+s+1,r+s,r+1,r+1\} + 
\{r+s+1,r+s+1,r+1,r\} \\ 
+ \{r+s+2,r+s,r+1,r\}+\{r+s+2,r+s+1,r,r\} 
\Rightarrow \\
\l[P\r]=\l[s-3/2,1/2,-1/2\r]+\l[s-1/2,3/2,-1/2\r] + \l[s-1/2,1/2,1/2\r] \\
+ \l[s+1/2,1/2,-1/2\r]+\l[s+1/2,3/2,1/2\r]+\l[s+3/2,1/2,1/2\r]\;.
\earr
\ee

\noindent $(iv.3):\;[\tilde{\mu}]=[1^3] \Rightarrow [\mu]
= [3,0,0,0] \Rightarrow \l[p\r]=\l[3/2,3/2,3/2\r]\;.$

\be
\barr{l}
\{\tilde{f}\}=\l\{4^r,3^3,2^{s-3}\r\}_{s \geq 3} +
\l\{4^r,3^2,2^{s-2},1\r\}_{s \geq 2} +
\l\{4^r,3,2^{s-1},1^2\r\}_{s \geq 1} + \l\{4^r,2^s,1^3\r\}_{s \geq 0} 
\Rightarrow \\
\{f\}=\{r+s,r+s,r+3,r\}+\{r+s+1,r+s,r+2,r\} \\ + \{r+s+2,r+s,r+1,r\} + 
\{r+s+3,r+s,r,r\} \Rightarrow \\
\l[P\r]=\l[s-3/2,3/2,-3/2\r]+\l[s-1/2,3/2,-1/2\r]+\l[s+1/2,3/2,1/2\r]\\
+ \l[s+3/2,3/2,3/2\r]\;. 
\earr
\ee

\noindent {\Large{$(v)\;$\bf{$v= 4$}}}

\noindent From Eq. (32) one has,
$$ 
\barr{l}
s=0,2,4,...,2k-2 \quad \mbox{ for } m=4k \\ 
s=1,3,5,...,2k-1 \quad \mbox{ for } m=4k+2\;. 
\earr
$$

\noindent $(v.1):\;[\tilde{\mu}]=[4] \Rightarrow [\mu]=[1,1,1,1] 
\Rightarrow [p]=[0,0,0]\;.$

\be
\barr{l}
\{\tilde{f}\}=\l\{4^{r+1},2^s\r\}_{s \geq 0}  \Rightarrow \\
\{f\}=\{r+s+1,r+s+1,r+1,r+1 \} \Rightarrow \\
\l[P\r]=\l[s,0,0\r]\;. 
\earr
\ee

\noindent $(v.2):\;[\tilde{\mu}]=[3,1] \Rightarrow [\mu]
= [2,1,1,0] \Rightarrow [p]=[1,1,0]\;.$
 
\be
\barr{l}
\{\tilde{f}\}=\l\{4^{r+1},3,2^{s-2},1\r\}_{s \geq 2} +
\l\{4^{r+1},2^s\r\}_{s \geq 1} +
\l\{4^{r+1},2^{s-1},1^2\r\}_{s \geq 1} \\
+ \l\{4^r,3^2,2^{s-1}\r\}_{s \geq 1} +
\l\{4^r,3,2^s,1\r\}_{s \geq 0}  \Rightarrow \\
\{f\}=\{r+s+1,r+s,r+2,r+1\}+\{r+s+1,r+s+1,r+1,r+1\} \\
+ \{r+s+2,r+s,r+1,r+1\}+ \{r+s+1,r+s+1,r+2,r\} \\
+\{r+s+2,r+s+1,r+1,r\} \Rightarrow \\
\l[P\r]=\l[s-1,1,0\r]+\l[s,0,0\r]+\l[s,1,1\r]+\l[s,1,-1\r]+\l[s+1,1,0\r]\;.
\earr
\ee

\noindent $(v.3):\;[\tilde{\mu}]=[2^2] \Rightarrow [\mu]
= [2,2,0,0] \Rightarrow [p]=[2,0,0]\;.$

\be
\barr{l}
\{\tilde{f}\}=\l\{4^{r+2},2^{s-2}\r\}_{s \geq 2} + 
\l\{4^{r+1},3,2^{s-2},1\r\}_{s \geq 2} + 
\l\{4^r,3^2,2^{s-2},1^2\r\}_{s \geq 2} \\
+ \l\{4^{r+1},2^s\r\}_{s \geq 1} + \l\{4^r,3,2^s,1\r\}_{s \geq 1} + 
\l\{4^r,2^{s+2}\r\}_{s \geq 0}  \Rightarrow \\
\{f\}=\{r+s,r+s,r+2,r+2\}+\{r+s+1,r+s,r+2,r+1\} \\
+ \{r+s+2,r+s,r+2,r\} + \{r+s+1,r+s+1,r+1,r+1\} \\
+ \{r+s+2,r+s+1,r+1,r\}+\{r+s+2,r+s+2,r,r\} \Rightarrow \\
\l[P\r]=\l[s-2,0,0\r]+\l[s-1,1,0\r]+\l[s,2,0\r]+\l[s,0,0\r]+\l[s+1,1,0\r]+
\l[s+2,0,0\r]\;.
\earr
\ee
Note that, for example $[P]=[2,0,0]$ appears for $\{\tilde{f}\}=\{4^r,2^2\}$ 
and $\{\tilde{f}\}=\{4^{r+1},2^2\}$, both for $m=4k$.

\noindent $(v.4):\;[\tilde{\mu}]=[2,1^2] \Rightarrow [\mu] =
[3,1,0,0] \Rightarrow [p]=[2,1,1]\;.$

\be
\barr{l}
\{\tilde{f}\}=\l\{4^{r+1},3,2^{s-2},1\r\}_{s \geq 2} +
\l\{4^r,3^2,2^{s-1}\r\}_{s \geq 2} + \l\{4^r,3^2,2^{s-2},1^2\r\}_{s \geq 2} 
+ \l\{4^{r+1},2^{s-1},1^2\r\}_{s \geq 1} \\
+ \l\{4^r,3,2^s,1\r\}_{s \geq 1} + \l\{4^r,3,2^{s-1},1^3\r\}_{s \geq 1} +
\l\{4^r,2^{s+1},1^2\r\}_{s \geq 0} \Rightarrow \\
\{f\}=\{r+s+1,r+s,r+2,r+1\} + \{r+s+1,r+s+1,r+2,r\} \\
+ \{r+s+2,r+s,r+2,r\} + \{r+s+2,r+s,r+1,r+1\} \\
+ \{r+s+2,r+s+1,r+1,r\} +\{r+s+3,r+s,r+1,r\} \\
+ \{r+s+3,r+s+1,r,r\} \Rightarrow \\
\l[P\r]=\l[s-1,1,0\r]+\l[s,1,-1\r]+\l[s,2,0\r]+\l[s,1,1\r]+\l[s+1,1,0\r]+
\l[s+1,2,1\r]\\
+\l[s+2,1,1\r]\;. 
\earr
\ee
Note that, for example $[P]=[2,1,1]$ appears for $\{\tilde{f}\} =
\{4^r,2,1^2\}$ and $\{\tilde{f}\}=\{4^{r+1},2,1^2\}$ both for $m=4k$.

\noindent $(v.5):\;[\tilde{\mu}]=[1^4] \Rightarrow [\mu] = [4,0,0,0]
\Rightarrow [p]=[2,2,2]\;.$

\be
\barr{l}
\{\tilde{f}\}= \l\{4^r,3^4,2^{s-4}\r\}_{s \geq 4} +
\l\{4^r,3^3,2^{s-3},1\r\}_{s \geq 3} +
\l\{4^r,3^2,2^{s-2},1^2\r\}_{s \geq 2} \\
+ \l\{4^r,3,2^{s-1},1^3\r\}_{s \geq 1} + \l\{4^r,2^s,1^4\r\}_{s \geq 0} 
\Rightarrow \\
\{f\}=\{r+s,r+s,r+4,r\}+\{r+s+1,r+s,r+3,r\}+\{r+s+2,r+s,r+2,r\} \\
+ \{r+s+3,r+s,r+1,r\}+\{r+s+4,r+s,r,r\} \Rightarrow \\
\l[P\r]=\l[s-2,2,-2\r]+\l[s-1,2,-1\r]+\l[s,2,0\r]+\l[s+1,2,1\r]
+\l[s+2,2,2\r]\;. 
\earr
\ee

\subsection{$O_{ST}(6) \supset O_S(3) \otimes O_T(3)$ reductions}

Eqs. (34)-(45) give the allowed $U_{ST}(4)$ or equivalently $O_{ST}(6)$ 
irreps (i.e. $\{f\}$ or $[P]$) that belong to the $O(8)$ seniority $v \leq
4$. With this, the remaining problem is to obtain the $O_{ST}(6) \supset
O_S(3) \otimes O_T(3)$ reductions. To solve this, we consider the equivalent
chain $SU_{ST}(4) \supset SU_S(2) \otimes SU_T(2)$. Firstly, given the
$U_{ST}(4)$ irrep $\{f\}=\{f_1, f_2, f_3, f_4\}$, the equivalent $SU_{ST}(4)$
irreps is $\{\lambda\} = \{\lambda_1,
\lambda_2,\lambda_3\}=\{f_1-f_4,f_2-f_4,f_3-f_4\}$. The general plethysm
problem to be solved is $([\spin]^\prime [\spin]^{\prime\prime}) \otimes
\{\lambda\}$. Now, using the simple result that a totally symmetric irrep
$\{\sigma\}$ of $U_{ST}(4)$ reduces to the irreps $\{\sigma_1,\sigma_2\}$ and
$\{\sigma_1,\sigma_2\}$, with $\sigma_1+\sigma_2=\sigma$, of $U_S(2)$ and
$U_T(2)$ in $U_{ST}(4) \supset U_S(2) \otimes U_T(2)$ and translating this 
to $SU_{ST}(4) \supset SU_S(2) \otimes SU_T(2)$, one has
\be
\{u\}_{SU(4)} \rightarrow (ST)=\l(\frac{u}{2}, \frac{u}{2}\r)
\oplus \l(\frac{u}{2}-1, \frac{u}{2}-1\r) \oplus \ldots \oplus
\l(\frac{\pi(u)}{2},\frac{\pi(u)}{2}\r)\;.
\ee
In Eq. (46) the symbol $\pi(r)$ is defined by,
\be
\pi(r)=0\;\;\;\mbox{if}\;\;\;r=\mbox{even},\;\;\;\;\;\;\pi(r)=1\;\;\;
\mbox{if}\;\;r=\mbox{odd}\;.
\ee
The reductions for a general $SU_{ST}(4)$ irrep 
$\{\lambda_1,\lambda_2,\lambda_3\}$ follow by expanding the Schur function
$\{\lambda_1,\lambda_2,\lambda_3\}$ in terms of totally symmetric Schur
functions and this is symbolically written as \cite{Ja-84}
\be
\{\lambda_1,\lambda_2,\lambda_3\}=\l|\barr{ccc} \{\lambda_1\} & 
\{\lambda_1+1\} &
\{\lambda_1+2\} \\ \{\lambda_2-1\} & \{\lambda_2\} & \{\lambda_2+1\} \\
\{\lambda_3-2\} & \{\lambda_3-1\} & \{\lambda_3\} \earr \r|
\ee
with $\{0\}=1$ and $\{-\lambda\}=0$. Now the basic quantity to compute is 
$([\spin]^\prime [\spin]^{\prime\prime}) \otimes \{u_1\} \{u_2\} \{u_3\}$ with
$\{u_i\}$'s totally symmetric, and it is nothing but the Kronecker product
of  $([\spin]^\prime  [\spin]^{\prime\prime}) \otimes \{u_1\}$,
$([\spin]^\prime  [\spin]^{\prime\prime}) \otimes \{u_2\}$ and 
$([\spin]^\prime 
[\spin]^{\prime\prime}) \otimes \{u_3\}$. This result together with Eqs. (46)
and (48) will completely solve the reduction problem and it is easy to
implement this procedure on a machine. It is seen easily from Eqs. (34) to
(37) that the $SU(4)$ irreps of interest for $v \leq 2$ are of the type
$\{s,s\}$, $\{s+1,s\}$, $\{s,s,1\} \sim \{s,s-1\}$, $\{s+2,s\}$, $\{s,s,2\}
\sim \{s,s-2\}$ and $\{s+1,s,1\}$. For these irreps, generating the
reductions to $(ST)$'s for $s \leq 10$, the following analytic formulas,
valid for any  $u$ are derived,
\be
\barr{l}
\{u,u\} \rightarrow (ST):\;S+T=u,u-2,\ldots, \pi(u) \nn8
\{u+1,u\} \rightarrow (ST):\;S+T=u+1,u,\ldots,1 \nn8
\{u+2,u\} \rightarrow (ST):\;\l\{\barr{l} (i)\;\; S+T= u,u-2,
\ldots, \pi(u) \\  
(ii)\;\;S+T=r,\; r=u+2,u+1,\ldots,2 \;\l\{\barr{l} S_{max}=r-1 \\ 
T_{max}=r-1 \earr\r.\earr\r. \nn8
\{u+1,u,1\} \rightarrow (ST):\;\l\{\barr{l} (i)\;\; S+T= u,u-1,\ldots,1
\\ (ii)\;\;S+T=r,\; r=u+1,u,\ldots,2 \;\l\{\barr{l} S_{max}=r-1 \\
T_{max}=r-1 \earr \r. \earr\r. \;. 
\earr
\ee

For Limit-I (i.e. for the chain (9)), as an example, the quantum numbers 
obtained using Eqs. (34), (36), (37) and (49) for $m=12$ are shown in Table I 
for $v=0$ and Table II for $v=2$. As seen from Eq. (19), all the $(ST)$ states
that belong to a $O(6)$ irrep $[P_1 P_2 P_3]$ are degenerate and the 
spacing between various $[P_1 P_2 P_3]$ multiplets for $v=0$ (then 
$[P_1 P_2 P_3]=[P,0,0]$), is $P(P+4)$ just as in the orbital space in the
$\gamma$-soft $O(6)$ limit of the interacting boson model (IBM) \cite{Ia-87}.
 
\section{Irrep reductions for the chain $U(4\Omega) \supset 
\l[ U(2\Omega) \supset Sp(2\Omega) \supset O(\Omega) \otimes SU_T(2)\r] 
\otimes SU_S(2)$}

For the chain $U(4\Omega) \supset \l[ U(2\Omega) \supset Sp(2\Omega) 
\supset O(\Omega) \otimes SU_T(2)\r] \otimes SU_S(2)$ (hereafter called
Limit-II) the branching rules are obtained by using in  much more detail the
plethysm method. Firstly, the reduction $U(2\Omega) \rightarrow Sp(2\Omega)$
is given by Eq.(80) of \cite{Wy-70},
\be
\barr{rcl}
\{\tilde{f}_S\}_{U(2\Omega)}  & = & \dis\sum_{\{\tilde{\mu}_S\}}\;
\dis\sum_{\{\tilde{\beta}\}} \;
\Gamma_{\{\tilde{\beta}\} \{\tilde{\mu}_S\} \{\tilde{f}_S}\}\;
\lan \tilde{\mu}_S \ran_{Sp(2\Omega)}\;, \nn8
\{\tilde{\beta}\} & = & \{0\}, \{1^2\}, \{1^4\}, \{2^2\}, \{1^6\}, 
\{2^21^2\}, \{3^2\}, \ldots
\earr
\ee
Note that the $\{\tilde{\beta}\}$ in (50) are the transpose of the
$\{\tilde{\delta}\}$ in the reduction $U(n) \rightarrow O(n)$ given by Eq.
(31). As pointed out in Section III.B, $\{\tilde{f}_S\}$ and 
$\lan \tilde{\mu}_S \ran$ are two columned irreps. Hence $\{\tilde{\beta}\}$
are of the form $\{2^{2\alpha_1}1^{2\alpha_2}\}$ with $\alpha_1$ and $\alpha_2$
being positive integers. Using this simplicity, all the reductions for
$U(2\Omega)$ irreps for $m \leq 16$ are generated using Eq. (50) and the
results are given in Table III only for $m \leq 12$ so that the table is not 
too long. 

Closely examining the results in Table III, the following band structures are
derived for $m \rightarrow S$ and $S \rightarrow (v_S,t_S)$ with
$t_S=0,\spin,1,\frac{3}{2}$, which are valid for any $m$,
\be
\begin{array}{l}
m=\mbox{even} \\
S=0,1,2,...,m/2 \\
t_S=0 \Rightarrow v_S=m-2S,m-2S-4,...,0 \mbox{ or } 2 \\
\\
t_S=1 \Rightarrow v_S=\left\{\begin{array}{l} (m-2S+2,m-2S,...,2)(
1-\delta_{S,0}) \\
m-2S-2,m-2S-6,...,2 \mbox{ or } 4 \end{array} \right. \\
\\ 
m=\mbox{odd } \\
S=1/2,3/2,...,m/2 \\
t_S=1/2 \Rightarrow v_S=m+1-2S,m-1-2S,...,1 \\
\\
t_S=3/2 \Rightarrow v_S=\left\{\begin{array}{l} (3,5,...,m+3-2S)(
1-\delta_{S,1/2})
\\
3,5,7,...,m-1-2S \end{array} \right. \;.
\end{array}
\ee
It is seen from Eq. (51) that for large $m$, there are two $v_S$ bands for
$t_S=1$ and $3/2$. For finding the branching rules for the 
$Sp(2\Omega) \rightarrow O(\Omega) \otimes SU_T(2)$ reductions, one first 
notes that, Eq. (128) of \cite{Wy-70} can be used (with $2\ell+1=\Omega$ 
in this equation). Then the irrep $\lan 1 \ran$ of $Sp(2\Omega)$ reduces 
into $O(\Omega) \otimes SU_T(2)$ irreps according to
\be
\lan 1 \ran= \l[1\r]^{\prime} \l[1/2\r]^{\prime \prime}\;. 
\ee
Then, by the theorem in Eq. (132) of \cite{Wy-70}, one obtains that the 
irreps of $O(\Omega)\otimes SU_T(2)$ contained in the irrep $<\tilde{\mu}_S>$ 
of $Sp(2\Omega)$ are those contained in the plethysm
$$
([1]^{\prime}[1/2]^{\prime \prime})\otimes <\tilde{\mu}_S>\;.
$$
Now, these irreps should be converted into Schur functions,
\be
\barr{l}
\l[1\r]^{\prime}=\{1\}^{\prime}\;, \\
\l[1/2\r]^{\prime \prime}=\{1\}^{\prime \prime} \;,\\
\lan \tilde{\mu}_S \ran = \{\tilde{\mu}_S \}+\dis\sum_{\{\rho\}} 
\left[\dis\sum_{\{\alpha\}}
(-1)^{r_{\alpha}/2} \Gamma_{\{\alpha\} \{\rho\} \{\tilde{\mu}_S\}} \right] 
\{\rho\}\;.
\earr
\ee
The last equation is Eq. (74) of \cite{Wy-70}. Also in (53), $\{\alpha\}$ 
are partitions of the form
$$
\left(\begin{array}{c} a \\ a+1 \end{array} \right), \;
\left(\begin{array}{cc} a & b \\ a+1 & b+1\end{array} \right),...
$$
in Frobenius notation and  $r_\alpha=\alpha_1+\alpha_2+\ldots$ is the degree
of $\{\alpha$\}. Note that these irreps $\{\alpha\}$ are the transposed of 
irreps $\{\gamma\}$ in Eq.(64) of \cite{Wy-70}. Now $([1]^{\prime}
[1/2]^{\prime \prime})\otimes <\tilde{\mu}_S>$ can be written as,
\be
\barr{l}
([1]^{\prime} [1/2]^{\prime \prime})\otimes \lan \tilde{\mu}_S \ran= \\
(\{1\}^{\prime} \{1\}^{\prime \prime})\otimes \left\{ \{\tilde{\mu}_S\} +
\dis\sum_{\{\rho\}} \left[\dis\sum_{\{\alpha\}}(-)^{r_{\alpha}/2}\Gamma_{
\{\alpha\} \{\rho\} \{\tilde{\mu}_S\}}
\right] \{\rho\} \right\} = \\
(\{1\}^{\prime} \{1\}^{\prime \prime})\otimes \{\tilde{\mu}_S\}+
\dis\sum_{\{\rho\}} \left[
\dis\sum_{\{\alpha\}}(-)^{r_{\alpha}/2}\Gamma_{\{\alpha\} \{\rho\} 
\{\tilde{\mu}_S\}}\right]
(\{1\}^{\prime}\{1\}^{\prime \prime})\otimes \{\rho\}\;.  
\earr
\ee
The fact that $\{\tilde{\mu}_S\}$ is a two-column irrep implies, 
that $\{\alpha\}$ and $\{\rho\}$ also have this property.  Then we will use 
the equation 
\be
\l\{2^a,1^b\r\}=\{1^{a+b}\}\{1^a\}-\{1^{a+b+1}\}\{1^{a-1}\}; \;\; a>1
\ee
to compute the plethysms that appear in Eq. (54). The net plethysm calculation 
then will be calculations of type
\be
(\{1\}^{\prime}\{1\}^{\prime \prime})\otimes \l\{1^p\r\}=
\dis\sum_{k_1+k_2=p} \l\{2^{k_2},1^{k_1-k_2}\r\}^{\prime}\; 
\l\{k_1,k_2\r\}^{\prime \prime}\;\;\mbox{for}\;\; Sp(2\Omega) \supset 
O(\Omega) \otimes SU_T(2)\;. 
\ee
After computing the plethysms that appear in Eq. (54) using Eqs. (55) 
and (56), the resulting Schur functions $\{\mu\}^{\prime}$ are converted into 
$O(\Omega)$ irreps using Eq.(69) of \cite{Wy-70}. Programming this procedure,
the reductions are obtained for $v_S=1-16$ with $v \leq 4$. These results for $v
\leq 3$ are given in Tables IV-X. To avoid making the paper too long, the 
tables for $v=4$ (i.e. for $[\tilde{\mu}] = [4]$, $[31]$, $[22]$, $[211]$ and
$[1111]$) are not given but they can be had upon request from the authors.

Results in Tables IV-X can be arranged, quite strikingly, into band structures
in isospin space and they are: 
\be
\begin{array}{l}
v=0,\;\;\left[\tilde{\mu}\right]=\left[0\right] \rightarrow <\tilde{\mu}_S>
=<2^r>: T=\pi(r),\pi(r)+2,
...,r\,,\;\;r \geq 0 \\
\\
v=1,\;\;\left[\tilde{\mu}\right]=\left[1\right] \rightarrow <\tilde{\mu}_S>
=<2^r1>:T=1/2,3/2,...,
r+1/2\,,\;\;r \geq 0 \\
\\
v=2,\;\;\left[\tilde{\mu}\right]=\left[2\right] \rightarrow <\tilde{\mu}_S>
=<2^r>: T=\left\{
\begin{array}{l} 1,2,3,...,r-1\,,\;\;r \geq 2 \\ \pi(r), \pi(r)+2,...,r\,,\;\;r \geq
1 \end{array} \right. \\
v=2,\;\;\left[\tilde{\mu}\right]=\left[2\right] \rightarrow <\tilde{\mu}_S>
=<2^r1^2>: T=\pi(r), 
\pi(r)+2,....,r\,,\;\;r \geq 0 \\
\\
v=2,\;\;\left[\tilde{\mu}\right]=\left[1^2\right] \rightarrow <\tilde{\mu}_S>
=<2^r>: T=1-\pi(r),
3-\pi(r),....,r-1\;,\;\;r \geq 1 \\
v=2,\;\;\left[\tilde{\mu}\right]=\left[1^2\right] \rightarrow <\tilde{\mu}_S>
=<2^r1^2>: T=\left\{
\begin{array}{l} 1,2,3,...,r\,,\;\;r \geq 1 \\ 1-\pi(r), 3-\pi(r),...,r+1\,,
\;\;r \geq 0 \end{array} \right. 
\\
v=3,\;\;\left[\tilde{\mu}\right]=\left[3\right] \rightarrow <\tilde{\mu}_S>
=<2^r1>:T=1/2,3/2,...,
r-1/2\,,\;\;r \geq 1 \\
\\
v=3,\;\;\left[\tilde{\mu}\right]=\left[21\right] \rightarrow <\tilde{\mu}_S>
=<2^r1>:T=\left\{\begin{array}{l} 
1/2,3/2,...,r+1/2\,,\;\;r \geq 1 \\
1/2,3/2,...,r-1/2\,,\;\;r \geq 2 \\
3/2,5/2,...,r-3/2\,,\;\;r \geq 3 \end{array} \right. \\
v=3,\;\;\left[\tilde{\mu}\right]=\left[21\right] \rightarrow <\tilde{\mu}_S>
=<2^r1^3>: T=1/2,3/2,...,
r+1/2\,,\;\;r \geq 0 \\
\\
v=3,\;\;\left[\tilde{\mu}\right]=\left[1^3\right] \rightarrow <\tilde{\mu}_S>
=<2^r1>: T=1/2,3/2,...,
r-1/2\,,\;\;r \geq 1 \\
v=3,\;\;\left[\tilde{\mu}\right]=\left[1^3\right] \rightarrow <\tilde{\mu}_S>
=<2^r1^3>: T=\left\{
\begin{array}{l} 3/2,5/2,...,r+3/2\,,\;\;r \geq 0 \\ 
1/2,3/2,...,r-1/2\,,\;\;r \geq 1 \end{array} \right.\;. \\
\end{array}
\ee
Combining Eqs. (51) and (57), for fixed $v$ and $[\tilde{\mu}]$ the 
reductions for $m \rightarrow S$, $S \rightarrow (v_S,t_S)$ and
$(v_S,t_S) \rightarrow T$ can be written down. Note that in Eqs. (57), $\lan
\tilde{\mu_S}\ran=\lan 2^{\mu_1} 1^{\mu_2} \ran \Rightarrow v_S=2\mu_1+\mu_2$
and $t_S=\mu_2/2$. Also it seen that for large $v_S$, the $T$'s form a band.

For Limit-II (i.e. for the chain (20)), as an example, the quantum numbers 
obtained using Eqs. (51) and (57) for $m=12$ are shown in Table I for $v=0$ 
and Table XI for $v=2$. Energy formula given by Eq. (28) and the results in 
Tables I and XI show that (as the energies are independent of isospin
$T$ and states with smaller $v_S$ will be lower in energy), the chain (20)
generates vibrational type spectra where $v_S/2$ can be viewed as the phonon
number. This is similar to the vibrational $U(5)$ limit (in orbital space)
of IBM \cite{Ia-87}. It is also important to remark that the total set of
$(ST)$ values for a given $v,[\tilde{\mu}]$ should be same for all the
three limits. This is verified by the results in Table I for $v=0$.
Similarly, this is verified for $v=2$ (and $[\tilde{\mu}]=[2]$ and $[1^2]$
separately) by the results in Tables II and XI (also by Table XII ahead).

\section{Irrep reductions for the chain $U(4\Omega) \supset 
\l[ U(2\Omega) \supset Sp(2\Omega) \supset O(\Omega) \otimes SU_S(2)\r] 
\otimes SU_T(2)$}

For the chain $U(4\Omega) \supset \l[ U(2\Omega) \supset Sp(2\Omega) 
\supset O(\Omega) \otimes SU_S(2)\r] \otimes SU_T(2)$ (hereafter called
Limit-III), all the irrep reductions can be obtained by just 
interchanging $S$ with $T$ in the
results in Section V. It is important to stress that in physical applications
it is useful to have band structures for  a given $S$. However what we obtain
via the results in Section V is  $m \rightarrow T$, $T \rightarrow v_T$  for
various allowed $t_T$, and $(v_T t_T) \rightarrow S$ for the $O(\Omega)$
irreps  $[\tilde{\mu}]=[0], [1], [2], [1^2], [3], [21], [1^3]$. Using these
results,  we have constructed tables for $m \rightarrow S$, $S \rightarrow
(v_T, t_T, T)^\alpha$ where $\alpha$ is the multiplicity.  Using these tables
the following band structures are derived for $v \leq 2$. For $v=0$,
\be
\barr{l}
m=\mbox{even},\;\;\left[\tilde{\mu}\right]=\left[0\right]\;: \\
S=0,1,2,...,m/2 \\
S \rightarrow t_T=0, \;\; v_T=2S,2S+4,...,m \mbox{ or } m-2 \\
v_T \rightarrow T=\pi(r),\pi(r)+2,...,r\,,\;\;r=(m-v_T)/2\;. 
\earr
\ee
For $v=1$,
\be
\barr{l}
m=\mbox{odd },\;\;\left[\tilde{\mu}\right]=\left[1\right]\;: \\ 
S=1/2,3/2,...,m/2 \\
S \rightarrow t_T=1/2, \;\;v_T=2S,2S+2,...,m \\
v_T \rightarrow T=1/2,3/2,\ldots,(m-v_T+1)/2\;.
\end{array}
\ee
For $v=2$,
\be
\begin{array}{l}
m=\mbox{even},\;\;\left[\tilde{\mu}\right]=\left[2\right]\;: \\
S=0,1,2,...,m/2 \\
S \rightarrow t_T=0, \;\; v_T=\left\{\begin{array}{l}
(2S,2S+2,\ldots,m)(1-\delta_{S,0}) \\
2S+4,2S+8,\ldots, m \mbox{ or } m-2 \end{array}\right. \\
v_T \rightarrow T=\pi(r),\pi(r)+2,...,r;\;\;r=(m-v_T)/2 \nn8
S \rightarrow t_T=1, \;\; v_T=2S+2,2S+6,\ldots, m \mbox{ or } m-2 \\
v_T \rightarrow T=\left\{\begin{array}{l} 1,2,3,...,r-1\,,\;\;r \geq 2 \\ 
\pi(r), \pi(r)+2,\ldots,r\,,\;\;r \geq 0 \end{array} \right\}\;;\;\;
r=(m-v_T+2)/2\;, 
\end{array}
\ee
$$
\begin{array}{l}
m=\mbox{even},\;\;\left[\tilde{\mu}\right]=\left[1^2\right]\;: \\
S=0,1,2,...,m/2; \\ 
S \rightarrow t_T=0, \;\; v_T=2S+2,2S+6,\ldots, m \mbox{ or } m-2 \\
v_T \rightarrow T=\pi(r),\pi(r)+2,...,r\,;\;\;r=(m-v_T)/2 
\earr
$$
\be
\begin{array}{l}
S \rightarrow t_T=1, \;\; v_T=\left\{\begin{array}{l}
(2S,2S+2,\ldots,m)(1-\delta_{S,0}) \\
2S+4,2S+8,\ldots, m \mbox{ or } m-2 \end{array}\right. \\
v_T \rightarrow T=\left\{\begin{array}{l} 1,2,3,...,r\,,\;\;r \geq 1 \\ 
1-\pi(r), 3-\pi(r),\ldots,r+1\,,\;\;r \geq 0 \end{array} \right\}
\,;\;\;r=(m-v_T)/2 \;. 
\end{array}
\ee
It is also possible to write down the formulas for $v=3$ but they are not
given here.

For Limit-III (i.e. for the chain (29)), as an example, the quantum numbers 
obtained using Eqs. (58), (60) and (61) for $m=12$ are shown in Table I  for
$v=0$ and Table XII for $v=2$. Energy formula given by Eq. (30) and the
results in Tables I and XII show that, as the energies have $T(T+1)$ 
dependence and states with smaller $v_T$ will be lower in energy, 
the chain (29) generates rotational spectra in isospace. This is similar to 
the rotational $SU(3)$ limit (in orbital space) of IBM \cite{Ia-87}.

\section{CONCLUSIONS}

In this article the long standing problem of classifying low-lying states
in  the $O(8)$ proton-neutron pairing model is solved. All the results are
derived for states with $O(8)$ seniority $v \leq 4$. Extensive use is made
of the recent developments \cite{Us} in the plethysm method in group
theory.  Band structures in isospace  are also derived in all the three
limits of the model for $v \leq 3$.  For Limit I (defined by Eq. (9)) they
are given by Eqs. (34)-(45) and (49). For Limit II (defined by Eq. (20))
they are given by Eqs. (51) and (57). Finally for Limit III (it is defined
by Eq. (29)) they are given by Eqs. (58)-(61). Some typical numerical
results are given for $m=12$ with $v=0$ and $2$ in Tables I, II, XI and XII.
From these tables and the energy formulas given by  Eqs. (19), (28) and
(30), it is seen that the three symmetry limits, strikingly, generate
vibrational, rotational and $\gamma$-soft like spectra (as in the three
limits of IBM \cite{Ia-87}) in isospace. 

Throughout this article the $O(\Omega) \supset O_L(3)$ branch is dropped. For
completeness we mention that the reductions of the irreps 
$[\tilde{\mu}]_{O(\Omega)}$, for $v \leq 4$, to $L$'s follow easily by
starting with the chain $U(\Omega)  \supset O(\Omega) \supset O_L(3)$ and the
basic association $\{1\}_{U(\Omega)} \rightarrow
(\ell_1,\ell_2,\ldots,\ell_r)_{O_L(3)}$ for nucleons occupying the orbits
$(\ell_1,\ell_2,\ldots,\ell_r)$; $\Omega=\sum_{i=1}^r\,(2\ell_i+1)$.  Now,
Eq. (48)  and the known methods \cite{Ko-old} for the reduction of symmetric
and  antisymmetric irreps of $U(\Omega)$ to $L$'s will give the reductions
$\{f\}_{U(\Omega)} \rightarrow L_{O(3)}$ where $\{f\}$'s are the irreps for
$m \leq 4$. These and Eq. (31) will give finally, $[\tilde{\mu}] \rightarrow
L$ reductions for $v \leq 4$. This procedure is easy to implement on a
machine but, as mentioned in Section I, the results here will depend on
$(\ell_1,\ell_2,\ldots,\ell_r)$.

With the classification of states available, the next step is to
develop Wigner-Racah algebra for the three chains of the $O(8)$ model (for
the $O_{ST}(6)$ chain some results are available for $v \leq 2$
\cite{Pa-69,He-85}). With this it is possible to probe the $O(8)$ model in
much more detail and apply it to heavy N $\sim$ Z nuclei but this is for
future.

\acknowledgements

The work presented in this paper has started as a result of participation by
one of the authors (VKBK) in the symposium `Symmetries in Science XIII' held
at Bregenz (Austria) during July 20-24, 2003.

\newpage
\begin{table}
\begin{center}
\caption{Quantum numbers in the symmetry limits I, II and III [
defined by the group chains (9), (20) and (29)] for $m=12$ and $v=0$. For the
definition of $s$, see Section IV.}
\vskip 0.25cm
\begin{tabular}{rrrl|rrl|rrl}
\hline
\multicolumn{4}{c}{Limit-I} & \multicolumn{3}{c}{Limit-II} &
\multicolumn{3}{c}{Limit-III} \\
\hline
$s$ & $[P]_{O(6)}$ & $\{f\}_{SU(4)}$ & $(ST)$ & $S$ & $v_S$ & $T$ & $S$ & 
$v_T$ & $T$ \\ 
\hline
$0$ &$[0]$ & $\{0\}$ & $(00)$ & $0$ & $0$ & $0$ & $0$ & $0$ & $0,2,4,6$ \\
$2$ & $[2]$ & $\{22\}$ & $(20)$,$(11)$,$(02),$ & & $4$ & $0,2$ & & 
$4$ & $0,2,4$ \\
& & & $(00)$ & & $8$ & $0,2,4$ & & $8$ & $0,2$ \\
$4$ & $[4]$ & $\{44\}$ & $(40),(31),(22),(13),$ & & $12$ & $0,2,4,6$ & & 
$12$ & $0$ \\
& & & $(04),(20),(11),(02),$ & $1$ & $2$ & $1$ & $1$ & $2$ & $1,3,5$ \\
& & & $(00)$ & & $6$ & $1,3$ & & $6$ & $1,3$ \\
$6$ & $[6]$ & $\{66\}$ & $(60),(51),(42),(33),$ & & $10$ & $1,3,5$ & & $10$ & $1$
\\
& & & $(24),(15),(06),$ & $2$ & $0$ & $0$ & $2$ & $4$ & $0,2,4$ \\
& & & $(40),(31),(22),(13),$ & & $4$ & $0,2$ & & $8$ & $0,2$ \\
& & & $(04),(20),(11),(02),$ & & $8$ & $0,2,4$ & & $12$ & $0$ \\
& & & $(00)$ & $3$ & $2$ & $1$ & $3$ & $6$ & $1,3$ \\
& & & & & $6$ & $1,3$ & & $10$ & $1$ \\
& & & & $4$ & $0$ & $0$ & $4$ & $8$ & $0,2$ \\
& & & & & $4$ & $0,2$ & & $12$ & $0$ \\
& & & & $5$ & $2$ & $1$ & $5$ & $10$ & $1$ \\
& & & & $6$ & $0$ & $0$ & $6$ & $12$ & $0$ \\ 
\hline
\end{tabular}
\end{center}
\end{table}
\newpage
\begin{table}
\begin{center}
\caption{Quantum numbers in the symmetry limit-I [
defined by the group chain (9)] for $m=12$ and $v=2$. For the
definition of $s$ see Section IV. In the table $(ST)^r$ indicates that the
multiplicity of the irreps $(ST)$ is $r$.}
\vskip 0.25cm
\begin{tabular}{rrrrl}
\hline
$\l[\tilde{\mu}\r]$ & $s$ & $[P]_{O(6)}$ & $\{f\}_{SU(4)}$ & $(ST)$  \\ 
\hline
$[2]$ & $1$ & $[0]$ & $\{0\}$ & $(00)$ \\
& & $[11]$ & $\{211\}$ & $(10),(01),(11)$ \\
& & $[2]$ & $\{22\}$ & $(20),(11),(02),(00)$ \\
& $3$ & $[2]$ & $\{22\}$ & same as above \\
& & $[31]$ & $\{431\}$ & $(30),(21)^2,(12)^2,(03),(20),(11)^2,(02),(10),
(01),$ \\
& & & & $(31),(22),(13)$ \\
& & $[4]$ & $\{44\}$ & $(40),(31),(22),(13),(04),(20),(11),(02),(00)$ \\
& $5$ & $[4]$ & $\{44\}$ & same as above \\
& & $[51]$ & $\{651\}$ & $(50),(41)^2,(32)^2,(23)^2,(14)^2,(05),(40),(31)^2,
(22)^2,(13)^2,(04),$ \\
& & & & $(30),(21)^2,(12)^2,(03),(20),(11)^2,(02),(10),(01),$ \\
& & & & $(51),(42),(33),(24),(15)$ \\
& & $[6]$ & $\{66\}$ & $(60),(51),(42),(33),(24),(15),(06),$ \\
& & & & $(40),(31),(22),(13),(04),(20),(11),(02),(00)$ \\
$[1^2]$ & $1$ & $[11]$ & $\{211\}$ & $(10),(01),(11)$ \\
& & $[211]$ & $\{31\}$ & $(10),(01),(21),(12),(11)$ \\
& $3$ & $[21-1]$ & $\{332\}$ & same as for $\{31\}$ \\
& & $[31]$ & $\{431\}$ & $(30),(21)^2,(12)^2,(03),(20),(11)^2,(02),
(10),(01),$ \\
& & & & $(31),(22),(13)$ \\
& & $[411]$ & $\{53\}$ & $(30),(21)^2,(12)^2,(03),(10),(01),$ \\
& & & & $(41),(32),(23),(14),(31),(22),(13),(11)$ \\
& $5$ & $[41-1]$ & $\{552\}$ & same as for $\{53\}$ \\
& & $[51]$ & $\{651\}$ & $(50),(41)^2,(32)^2,(23)^2,(14)^2,(05),(40),(31)^2,
(22)^2,(13)^2,(04),$ \\
& & & & $(30),(21)^2,(12)^2,(03),(20),(11)^2,(02),(10),(01),$ \\
& & & & $(51),(42),(33),(24),(15)$ \\
& & $[611]$ & $\{75\}$ &
$(61),(52),(43),(34),(25),(16),(51),(42),(33),(24),(15)$ \\
& & & & $(50),(41)^2,(32)^2,(23)^2,(14)^2,(05),$ \\
& & & & $(31),(22),(13),(30),(21)^2,(12)^2,(03),(11),(10),(01)$ \\
\hline
\end{tabular}
\end{center}
\end{table}
\newpage
\begin{table}
\begin{center}
\caption{$U(2\Omega) \supset Sp(2\Omega)$ reductions for two columned
$U(2\Omega)$ irreps for $m=0-12$.}
\vskip 0.25cm
\begin{tabular}{rcl}
\hline
$\{0\}$ & $=$ & $<0>$ \\
$\{1\}$ & $=$ & $<1>$ \\
$\{2\}$ & $=$ & $<2>$ \\
$\{1^{2}\}$ & $=$ & $<1^{2}>+<0>$ \\
$\{21\}$ & $=$ & $<21>+<1>$ \\
$\{1^{3}\}$ & $=$ & $<1^{3}>+<1>$ \\
$\{2^{2}\}$ & $=$ & $<2^{2}>+<1^{2}>+<0>$ \\
$\{21^{2}\}$ & $=$ & $<21^{2}>+<2>+<1^{2}>$ \\
$\{1^{4}\}$ & $=$ & $<1^{4}>+<1^{2}>+<0>$ \\
$\{2^{2}1\}$ & $=$ & $<2^{2}1>+<21>+<1^{3}>+<1>$ \\
$\{21^{3}\}$ & $=$ & $<21^{3}>+<21>+<1^{3}>+<1>$ \\
$\{1^{5}\}$ & $=$ & $<1^{5}>+<1^{3}>+<1>$ \\
$\{2^{3}\}$ & $=$ & $<2^{3}>+<21^{2}>+<2>$ \\
$\{2^{2}1^{2}\}$ & $= $ & $<2^{2}1^{2}>+<2^{2}>+<21^{2}>+<1^{4}>+2<1^{2}>
+<0>$ \\
$\{21^{4}\}$ & $=$ & $<21^{4}>+<21^{2}>+<1^{4}>+<2>+<1^{2}>$ \\
$\{1^{6}\}$ & $=$ & $<1^{6}>+<1^{4}>+<1^{2}>+<0>$ \\
$\{2^{3}1\}$ & $=$ & $<2^{3}1>+<2^{2}1>+<21^{3}>+<1^{3}>+<21>+<1>$ \\
$\{2^{2}1^{3}\}$ & $=$ & $<2^{2}1^{3}>+<2^{2}1>+<21^{3}>+<1^{5}>+<21>+
2<1^{3}>+<1>$ \\
$\{21^{5}\}$ & $=$ & $<21^{5}>+<21^{3}>+<1^{5}>+<21>+<1^{3}>+<1>$ \\
$\{1^{7}\}$ & $=$ & $<1^{7}>+<1^{5}>+<1^{3}>+<1>$ \\ 
\hline
\end{tabular}
\end{center}
\end{table}
\newpage
\begin{center}
Table III (Cont'd)
\begin{tabular}{rcl}
\hline
$\{2^{4}\}$ & $=$ & $<2^{4}>+<2^{2}1^{2}>+<1^{4}>+<2^{2}>+<1^{2}>+<0>$ \\
$\{2^{3}1^{2}\}$ & $=$ & $<2^{3}1^{2}>+<2^{3}>+<2^{2}1^{2}>+<21^{4}>+
2<21^{2}>+<1^{4}>$ \\
& $+$ & $<2>+<1^{2}>$ \\
$\{2^{2}1^{4}\}$ & $=$ & $<2^{2}1^{4}>+<2^{2}1^{2}>+<21^{4}>+<1^{6}>+<2^{2}>
+<21^{2}>$ \\
& $+$ & $2<1^{4}>+2<1^{2}>+<0>$ \\
$\{21^{6}\}$ & $=$ & $<21^{6}>+<21^{4}>+<1^{6}>+<21^{2}>+<1^{4}>+<2>+<1^{2}>$ \\
$\{1^{8}\}$ & $=$ & $<1^{8}>+<1^{6}>+<1^{4}>+<1^{2}>+<0>$ \\
$\{2^{4}1\}$ & $=$ & $<2^{4}1>+<2^{3}1>+<2^{2}1^{3}>+<21^{3}>+<1^{5}>+<2^{2}1>$
\\
& $+$ & $<21>+<1^{3}>+<1>$ \\
$\{2^{3}1^{3}\}$ & $=$ & $<2^{3}1^{3}>+<2^{3}1>+<2^{2}1^{3}>+<21^{5}>+<2^{2}1>
+2<21^{3}>$ \\
& $+$ & $<1^{5}>+2<1^{3}>+<21>+<1>$ \\
$\{2^{2}1^{5}\}$ & $=$ & $<2^{2}1^{5}>+<2^{2}1^{3}>+<21^{5}>+<1^{7}>+<2^{2}1>+
<21^{3}>$ \\
& $+$ & $2<1^{5}>+<21>+2<1^{3}>+<1>$ \\
$\{21^{7}\}$ & $=$ & $<21^{7}>+<21^{5}>+<1^{7}>+<21^{3}>+<1^{5}>+<21>
+<1^{3}>$ \\
& $+$ & $<1>$ \\
$\{1^{9}\}$ & $=$ & $<1^{9}>+<1^{7}>+<1^{5}>+<1^{3}>+<1>$ \\
$\{2^{5}\}$ & $=$ & $<2^{5}>+<2^{3}1^{2}>+<21^{4}>+<2^{3}>+<21^{2}>+<2>$ \\
$\{2^{4}1^{2}\}$ & $=$ & $<2^{4}1^{2}>+<2^{4}>+<2^{3}1^{2}>+<2^{2}1^{4}>
+2<2^{2}1^{2}>+<21^{4}>$ \\
& $+$ & $<1^{6}>+2<1^{4}>+<2^{2}>+<21^{2}>+2<1^{2}>+<0>$ \\
$\{2^{3}1^{4}\}$ & $=$ & $<2^{3}1^{4}>+<2^{3}1^{2}>+<2^{2}1^{4}>+<21^{6}>
+<2^{3}>+<2^{2}1^{2}>$ \\
& $+$ & $2<21^{4}>+<1^{6}>+2<21^{2}>+2<1^{4}>+<2>+<1^{2}>$ \\
$\{2^{2}1^{6}\}$ & $=$ & $<2^{2}1^{6}>+<2^{2}1^{4}>+<21^{6}>+<1^{8}>
+<2^{2}1^{2}>+<21^{4}>$ \\
& $+$ & $2<1^{6}>+<2^{2}>+<21^{2}>+2<1^{4}>+2<1^{2}>+<0>$ \\
\hline
\end{tabular}
\end{center}
\newpage
\begin{center}
Table III (Cont'd)
\begin{tabular}{rcl}
\hline
$\{21^{8}\}$ & $=$ & $<21^{8}>+<21^{6}>+<1^{8}>+<21^{4}>+<1^{6}>+<21^{2}>
+<1^{4}>$ \\
& $+$ & $<2>+<1^{2}>$ \\
$\{1^{10}\}$ & $=$ & $<1^{10}>+<1^{8}>+<1^{6}>+<1^{4}>+<1^{2}>+<0>$ \\
$\{2^{5}1\}$ & $=$ & $<2^{5}1>+<2^{4}1>+<2^{3}1^{3}>+<2^{2}1^{3}>+<21^{5}>
+<2^{3}1>$ \\
& $+$ & $<1^{5}>+<2^{2}1>+<21^{3}>+<1^{3}>+<21>+<1>$ \\
$\{2^{4}1^{3}\}$ & $=$ & $<2^{4}1^{3}>+<2^{4}1>+<2^{3}1^{3}>+<2^{2}1^{5}>
+<2^{3}1>+2<2^{2}1^{3}>$ \\
& $+$ & $<21^{5}>+<1^{7}>+2<21^{3}>+2<1^{5}>+<2^{2}1>+<21>$ \\
& $+$ & $2<1^{3}>+<1>$ \\
$\{2^{3}1^{5}\}$ & $=$ & $<2^{3}1^{5}>+<2^{3}1^{3}>+<2^{2}1^{5}>+<21^{7}>
+<2^{3}1>+<2^{2}1^{3}>$ \\
& $+$ & $2<21^{5}>+<1^{7}>+<2^{2}1>+2<21^{3}>+2<1^{5}>+2<1^{3}>$ \\
& $+$ & $<21>+<1>$ \\
$\{2^{2}1^{7}\}$ & $=$ & $<2^{2}1^{7}>+<2^{2}1^{5}>+<21^{7}>+<1^{9}>
+<2^{2}1^{3}>+<21^{5}>$ \\
& $+$ & $2<1^{7}>+<2^{2}1>+<21^{3}>+2<1^{5}>+<21>+2<1^{3}>+<1>$ \\
$\{21^{9}\}$ & $=$ & $<21^{9}>+<21^{7}>+<1^{9}>+<21^{5}>+<1^{7}>+<21^{3}>$ \\
& $+$ & $<1^{5}>+<21>+<1^{3}>+<1>$ \\
$\{1^{11}\}$ & $=$ & $<1^{11}>+<1^{9}>+<1^{7}>+<1^{5}>+<1^{3}>+<1>$ \\
$\{2^{6}\}$ & $=$ & $<2^{6}>+<2^{4}1^{2}>+<2^{2}1^{4}>+<2^{4}>+<1^{6}>
+<2^{2}1^{2}>$ \\
& $+$ & $<1^{4}>+<2^{2}>+<1^{2}>+<0>$ \\
$\{2^{5}1^{2}\}$ & $=$ & $<2^{5}1^{2}>+<2^{5}>+<2^{4}1^{2}>+<2^{3}1^{4}>
+2<2^{3}1^{2}>+<2^{2}1^{4}>$ \\
& $+$ & $<21^{6}>+2<21^{4}>+<1^{6}>+<2^{3}>+<2^{2}1^{2}>+2<21^{2}>+<1^{4}>$ \\
& $+$ & $<2>+<1^{2}>$ \\
$\{2^{4}1^{4}\}$ & $=$ & $<2^{4}1^{4}>+<2^{4}1^{2}>+<2^{3}1^{4}>+<2^{2}1^{6}>
+<2^{4}>+<2^{3}1^{2}>$ \\
& $+$ & $2<2^{2}1^{4}>+<21^{6}>+<1^{8}>+2<2^{2}1^{2}>+2<21^{4}>+2<1^{6}>$ \\
& $+$ & $3<1^{4}>+<2^{2}>+<21^{2}>+2<1^{2}>+<0>$ \\
\hline
\end{tabular}
\end{center}
\newpage
\begin{center}
Table III (Cont'd)
\begin{tabular}{rcl}
\hline
$\{2^{3}1^{6}\}$ & $=$ & $<2^{3}1^{6}>+<2^{3}1^{4}>+<2^{2}1^{6}>+<21^{8}>
+<2^{3}1^{2}>+<2^{2}1^{4}>$ \\
& $+$ & $2<21^{6}>+<1^{8}>+<2^{3}>+<2^{2}1^{2}>+2<21^{4}>+2<1^{6}>$ \\
& $+$ & $2<21^{2}>+2<1^{4}>+<2>+<1^{2}>$ \\
$\{2^{2}1^{8}\}$ & $=$ & $<2^{2}1^{8}>+<2^{2}1^{6}>+<21^{8}>+<1^{10}>
+<2^{2}1^{4}>+<21^{6}>$ \\
& $+$ & $2<1^{8}>+<2^{2}1^{2}>+<21^{4}>+2<1^{6}>+<2^{2}>+<21^{2}>$ \\
& $+$ & $2<1^{4}>+2<1^{2}>+<0>$ \\
$\{21^{10}\}$ & $=$ & $<21^{10}>+<21^{8}>+<1^{10}>+<21^{6}>+<1^{8}>
+<21^{4}>+<1^{6}>$ \\
& $+$ & $<21^{2}>+<1^{4}>+<2>+<1^{2}>$ \\
$\{1^{12}\}$ & $=$ & $<1^{12}>+<1^{10}>+<1^{8}>+<1^{6}>+<1^{4}>+<1^{2}>+<0>$ \\ 
\hline
\end{tabular}
\end{center}
\newpage
\begin{table}
\begin{center}
\caption{ $Sp(2\Omega)$ irreps $<\tilde{\mu}_S>$ that contain the $O(\Omega)$ 
irrep $[0]$, with multiplicities and associated $(2T+1)$ values given in 
column 3.}
\vskip 0.25cm
\begin{tabular}{ccr} \hline
       &             &   $[\tilde{\mu}]=[0]$       \\ \hline
 $v_S$   & $<\tilde{\mu}_S>$ &   $\alpha(2T+1)$       \\ \hline
  0    & $<0>$       &   1(1)                 \\
  2    & $<2>$       &   1(3)                 \\ 
  4    & $<2^2>$     &   1(5) 1(1)            \\ 
  6    & $<2^3>$     &   1(7) 1(3)            \\ 
  8    & $<2^4>$     &   1(9) 1(5) 1(1)       \\ 
 10    & $<2^5>$     &   1(11) 1(7) 1(3)      \\  
 12    & $<2^6>$     &   1(13) 1(9) 1(5) 1(1) \\  
 14    & $<2^7>$     &   1(15) 1(11) 1(7) 1(3) \\ 
 16    & $<2^8>$     &   1(17) 1(13) 1(9) 1(5) 1(1) \\ 
\hline
\end{tabular}
\end{center}
\end{table}

\vskip 1cm

\begin{table}
\begin{center}
\caption{ $Sp(2\Omega)$ irreps $<\tilde{\mu}_S>$ that contain the $O(\Omega)$ 
irrep $[1]$, with multiplicities and associated $(2T+1)$ values given in 
column 3.}
\vskip 0.25cm
\begin{tabular}{ccr} 
\hline
       &             & $[\tilde{\mu}]=[1]$                  \\ \hline
 $v_S$   & $<\tilde{\mu}_S>$ & $\alpha(2T+1)$                  \\ \hline
  1    & $<1>$       & 1(2)                            \\ 
  3    & $<21>$      & 1(4) 1(2)                       \\ 
  5    & $<2^21>$    & 1(6) 1(4) 1(2)                  \\
  7    & $<2^31>$    & 1(8) 1(6) 1(4) 1(2)             \\
  9    & $<2^41>$    & 1(10) 1(8) 1(6) 1(4) 1(2)       \\ 
 11    & $<2^51>$    & 1(12) 1(10) 1(8) 1(6) 1(4) 1(2) \\
 13    & $<2^61>$    & 1(14) 1(12) 1(10) 1(8) 1(6) 1(4) 1(2) \\ 
 15    & $<2^71>$    & 1(16) 1(14) 1(12) 1(10) 1(8) 1(6) 1(4) 1(2) \\ 
\hline
\end{tabular}
\end{center}
\end{table}
\newpage
\begin{table}
\begin{center}
\caption{ $Sp(2\Omega)$ irreps $<\tilde{\mu}_S>$ that contain the $O(\Omega)$ 
irrep $[2]$, with multiplicity and $2T+1$ values listed in column 3.}
\vskip 0.25cm
\begin{tabular}{ccr}\hline 
     &             &   $[\tilde{\mu}]=[2]$   \\ \hline
 $v_S$ & $<\tilde{\mu}_S>$ &   $\alpha(2T+1)$   \\ \hline
  2  & $<2>$       & 1(3)               \\ 
     & $<1^2>$     & 1(1)               \\ 
  4  & $<2^2>$     & 1(5) 1(3) 1(1)     \\
     & $<21^2>$    & 1(3)               \\ 
  6  & $<2^3>$     & 1(7) 1(5) 2(3)     \\
     & $<2^21^2>$  & 1(5) 1(1)          \\ 
  8  & $<2^4>$     & 1(9) 1(7) 2(5) 1( 3) 1(1)    \\
     & $<2^31^2>$  & 1(7) 1(3)          \\ 
  10 & $<2^5>$     & 1(11) 1(9) 2(7) 1( 5) 2(3)  \\
     & $<2^41^2>$  & 1(9) 1(5) 1(1)     \\ 
  12 & $<2^6>$     & 1(13) 1(11) 2(9) 1(7) 2(5) 1(3) 1(1)  \\  
     & $<2^51^2>$  & 1(11) 1(7) 1(3)    \\ 
  14 & $<2^7>$     & 1(15) 1(13) 2(11) 1(9) 2(7) 1(5) 2(3)  \\
     & $<2^61^2>$  & 1(13) 1(9) 1(5) 1(1) \\ 
  16 & $<2^8>$     & 1(17) 1(15) 2(13) 1(11) 2(9) 1(7) 2(5) 1(3) 1(1) \\
     & $<2^71^2>$  & 1(15) 1(11) 1(7) 1(3) \\ \hline
\end{tabular}
\end{center}
\end{table}
\newpage
\begin{table}
\begin{center}
\caption{ $Sp(2\Omega)$ irreps $<\tilde{\mu}_S>$ that contain the $O(\Omega)$ 
irrep $[1^2]$ with multiplicity and $2T+1$ values listed in columns 3.}
\vskip 0.25cm
\begin{tabular}{ccr}\hline 
     &             & $[\tilde{\mu}]=[1^2]$  \\ \hline
 $v_S$ & $<\tilde{\mu}_S>$ & $\alpha(2T+1)$    \\ \hline
  2  & $<2>$       & 1(1)               \\ 
     & $<1^2>$     & 1(3)               \\ 
  4  & $<2^2>$     & 1(3)               \\
     & $<21^2>$    & 1(5) 1(3) 1(1)     \\ 
  6  & $<2^3>$     & 1(5) 1(1)          \\
     & $<2^21^2>$  & 1(7) 1(5) 2(3)     \\ 
  8  & $<2^4>$     & 1(7) 1(3)   \\
     & $<2^31^2>$  & 1(9) 1(7) 2(5) 1(3) 1(1) \\ 
  10 & $<2^5>$     & 1(9) 1(5) 1(1) \\
     & $<2^41^2>$  & 1(11) 1(9) 2(7) 1(5) 2(3) \\ 
  12 & $<2^6>$     & 1(11) 1(7) 1(3) \\  
     & $<2^51^2>$  & 1(13) 1(11) 2(9) 1(7) 2(5) 1(3) 1(1)   \\ 
  14 & $<2^7>$     & 1(13) 1(9) 1(5) 1(1) \\
     & $<2^61^2>$  & 1(15) 1(13) 2(11) 1(9) 2(7) 1(5) 2(3)  \\ 
  16 & $<2^8>$     & 1(15) 1(11) 1(7) 1(3) \\
     & $<2^71^2>$  & 1(17) 1(15) 2(13) 1(11) 2(9) 1(7) 2(5) 1(3) 1(1)
 \\ \hline
\end{tabular}
\end{center}
\end{table}

\newpage

\begin{table}
\begin{center}
\caption{ $Sp(2\Omega)$ irreps $<\tilde{\mu}_S>$ that contain the $O(\Omega)$ 
irrep $[3]$, with multiplicities and associated $(2T+1)$ values given in 
column 3.}
\vskip 0.25cm
\begin{tabular}{ccr} \hline
     &             & $[\tilde{\mu}]=[3]$            \\ \hline
 $v_S$ & $<\tilde{\mu}_S>$ & $\alpha(2T+1)$            \\ \hline
  3  & $<21>$      & 1(2)                 \\ 
  5  & $<2^21>$    & 1(4) 1(2)                 \\ 
  7  & $<2^31>$    & 1(6) 1(4) 1(2)            \\ 
  9  & $<2^41>$    & 1(8) 1(6) 1(4) 1(2)       \\ 
 11  & $<2^51>$    & 1(10) 1(8) 1(6) 1(4) 1(2) \\ 
 13  & $<2^61>$    & 1(12) 1(10) 1(8) 1(6) 1(4) 1(2)  \\ 
 15  & $<2^71>$    & 1(14) 1(12) 1(10) 1(8) 1(6) 1(4) 1(2)  \\ \hline
\end{tabular}
\end{center}
\end{table}

\vskip 1cm

\begin{table}
\begin{center}
\caption{ $Sp(2\Omega)$ irreps $<\tilde{\mu}_S>$ that contain the $O(\Omega)$ 
irrep $[21]$, with multiplicities and associated $(2T+1)$ values given in 
column 3.}
\vskip 0.25cm
\begin{tabular}{ccr} \hline
       &             & $[\tilde{\mu}]=[21]$                 \\ \hline
 $v_S$   & $<\tilde{\mu}_S>$ & $\alpha(2T+1)$                  \\ \hline
  3    & $<21>$      & 1(4) 1(2)                  \\ 
       & $<1^3>$     & 1(2)                       \\ 
  5    & $<2^21>$    & 1(6) 2(4) 2(2)                  \\ 
       & $<21^3>$    & 1(4) 1(2)                       \\ 
  7    & $<2^31>$    & 1(8) 2(6) 3(4) 2(2)             \\
       & $<2^21^3>$  & 1(6) 1(4) 1(2)                  \\ 
  9    & $<2^41>$    & 1(10) 2(8) 3(6) 3(4) 2(2)       \\
       & $<2^31^3>$  & 1(8) 1(6) 1(4) 1(2)             \\ 
 11    & $<2^51>$    & 1(12) 2(10) 3(8) 3(6) 3(4) 2(2) \\
       & $<2^41^3>$  & 1(10) 1(8) 1(6) 1(4) 1(2)       \\ 
 13    & $<2^61>$    & (14) 2(12) 3(10) 3(8) 3(6) 3(4) 2(2) \\
       & $<2^51^3>$  & 1(12) 1(10) 1(8) 1(6) 1(4) 1(2) \\ 
 15    & $<2^71^1>$  & 1(16) 2(14) 3(12) 3(10) 3(8) 3(6) 3(4) 2(2) \\
       & $<2^61^3>$  & 1(14) 1(12) 1(10) 1(8) 1(6) 1(4) 1(2) \\ \hline
\end{tabular}
\end{center}
\end{table}

\newpage

\begin{table}
\begin{center}
\caption{ $Sp(2\Omega)$ irreps $<\tilde{\mu}_S>$ that contain the $O(\Omega)$ 
irrep $[1^3]$, with multiplicities and associated $(2T+1)$ values given in 
column 3.}
\vskip 0.25cm
\begin{tabular}{ccr} \hline
     &             & $[\tilde{\mu}]=[1^3]$                \\ \hline
 $v_S$ & $<\tilde{\mu}_S>$ & $\alpha(2T+1)$                  \\ \hline
  3  & $<21>$      & 1(2)                       \\
     & $<1^3>$     & 1(4)                   \\ 
  5  & $<2^21>$    & 1(4) 1(2)                       \\
     & $<21^3>$    & 1(6) 1(4) 1(2)                  \\ 
  7  & $<2^31>$    & 1(6) 1(4) 1(2)                  \\ 
     & $<2^21^3>$  & 1(8) 1(6) 2(4) 1(2)             \\ 
  9  & $<2^41>$    & 1(8) 1(6) 1(4) 1(2)             \\ 
     & $<2^31^3>$  & 1(10) 1(8) 2(6) 2(4) 1(2)       \\ 
 11  & $<2^51>$    & 1(10) 1(8) 1(6) 1(4) 1(2)       \\
     & $<2^41^3>$  & 1(12) 1(10) 2(8) 2(6) 2(4) 1(2) \\ 
 13  & $<2^61>$    & 1(12) 1(10) 1(8) 1(6) 1(4) 1(2) \\
     & $<2^51^3>$  & 1(14) 1(12) 2(10) 2(8) 2(6) 2(4) 1(2)  \\ 
 15  & $<2^71>$    & 1(14) 1(12) 1(10) 1(8) 1(6) 1(4) 1(2)  \\
     & $<2^61^3>$  & 1(16) 1(14) 2(12) 2(10) 2(8) 2(6) 2(4) 1(2) \\ \hline
\end{tabular}
\end{center}
\end{table}
\newpage
\begin{table}
\begin{center}
\caption{Quantum numbers in the symmetry limit-II [defined by the group 
chain (20)] for $m=12$ and $v=2$. In the last two columns
of the table, different $T$ bands are separated by a semicolon. In the table
$v_S^r$ indicates that the multiplicity of the irrep $(v_S,t_S)$ is $r$.}
\vskip 0.25cm
\begin{tabular}{rrrl|l}
\hline
& & & $\l[\tilde{\mu}\r]=[2]$ & $\l[\tilde{\mu}\r]=[1^2]$ \\
\hline
$t_S$ & $S$ & $v_S$ & $T$ & $T$ \\ 
\hline
$0$ & $0$ & $4$ & $1;0,2$ & $1$ \\
& & $8$ & $1,2,3;0,2,4$ & $1,3$ \\
& & $12$ & $1,2,3,4,5;0,2,4,6$ & $1,3,5$ \\
& $1$ & $2$ & $1$ & $0$ \\
& & $6$ & $1,2;1,3$ & $0,2$ \\
& & $10$ & $1,2,3,4;1,3,5$ & $0,2,4$ \\
& $2$ & $4$ & $1;0,2$ & $1$ \\
& & $8$ & $1,2,3;0,2,4$ & $1,3$ \\
& $3$ & $2$ & $1$ & $0$ \\
& & $6$ & $1,2;1,3$ & $0,2$ \\
& $4$ & $4$ & $1;0,2$ & $1$ \\
& $5$ & $2$ & $1$ & $0$ \\
\hline
$1$ & $0$ & $2$ & $0$ & $1$ \\
& & $6$ & $0,2$ & $1,2;1,3$ \\
& & $10$ & $0,2,4$ & $1,2,3,4;1,3,5$ \\
& $1$ & $2$ & $0$ & $1$ \\
& & $4^2$ & $1$ & $1;0,2$ \\
& & $6$ & $0,2$ & $1,2;1,3$ \\
& & $8^2$ & $1,3$ & $1,2,3;0,2,4$ \\
& & $10$ & $0,2,4$ & $1,2,3,4;1,3,5$ \\
& & $12$ & $1,3,5$ & $1,2,3,4,5;0,2,4,6$ \\
& $2$ & $2^2$ & $0$ & $1$ \\
& & $4$ & $1$ & $1;0,2$ \\
& & $6^2$ & $0,2$ & $1,2;1,3$ \\
& & $8$ & $1,3$ & $1,2,3;0,2,4$ \\
& & $10$ & $0,2,4$ & $1,2,3,4;1,3,5$ \\
& $3$ & $2$ & $0$ & $1$ \\
& & $4^2$ & $1$ & $1;0,2$ \\
& & $6$ & $0,2$ & $1,2;1,3$ \\
& & $8$ & $1,3$ & $1,2,3;0,2,4$ \\
& $4$ & $2^2$ & $0$ & $1$ \\
& & $4$ & $1$ & $1;0,2$ \\
& & $6$ & $0,2$ & $1,2;1,3$ \\
& $5$ & $2$ & $0$ & $1$ \\
& & $4$ & $1$ & $1;0,2$ \\
& $6$ & $2$ & $0$ & $1$ \\
\hline
\end{tabular}
\end{center}
\end{table}
\newpage
\begin{table}
\begin{center}
\caption{Quantum numbers in the symmetry limit-III [defined by the group 
chain (29)] for $m=12$ and $v=2$. For other details see Table XI.}
\vskip 0.25cm
\begin{tabular}{rrl|l}
\hline
& & $t_T=0,\;\;\l[\tilde{\mu}\r]=[2]$ & $t_T=1,\;\;\l[\tilde{\mu}\r]=[1^2]$ \\
\hline
$S$ & $v_T$ & $T$ & $T$ \\ 
\hline
$0$ & $4$ & $0,2,4$ & $1,2,3,4;1,3,5$ \\
& $8$ & $0,2$ & $1,2;1,3$ \\
& $12$ & $0$ & $1$ \\
$1$ & $2$ & $1,3,5$ & $1,2,3,4,5;0,2,4,6$ \\
& $4$ & $0,2,4$ & $1,2,3,4;1,3,5$ \\
& $6^2$ & $1,3$ & $1,2,3;0,2,4$ \\
& $8$ & $0,2$ & $1,2;1,3$ \\
& $10^2$ & $1$ & $1;0,2$ \\
& $12$ & $0$ & $1$ \\
$2$ & $4$ & $0,2,4$ & $1,2,3,4;1,3,5$ \\
& $6$ & $1,3$ & $1,2,3;0,2,4$ \\
& $8^2$ & $0,2$ & $1,2;1,3$ \\
& $10$ & $1$ & $1;0,2$ \\
& $12^2$ & $0$ & $1$ \\
$3$ & $6$ & $1,3$ & $1,2,3;0,2,4$ \\
& $8$ & $0,2$ & $1,2;1,3$ \\
& $10^2$ & $1$ & $1;0,2$ \\
& $12$ & $0$ & $1$ \\
$4$ & $8$ & $0,2$ & $1,2;1,3$ \\
& $10$ & $1$ & $1;0,2$ \\
& $12^2$ & $0$ & $1$ \\
$5$ & $10$ & $1$ & $1;0,2$ \\
& $12$ & $0$ & $1$ \\
$6$ & $12$ & $0$ & $1$ \\
\hline
& & $t_T=0,\;\;\l[\tilde{\mu}\r]=[1^2]$ & $t_T=1,\;\;\l[\tilde{\mu}\r]=[2]$ \\
\hline
$0$ & $2$ & $1,3,5$ & $1,2,3,4,5;0,2,4,6$ \\
& $6$ & $1,3$ & $1,2,3;0,2,4$ \\
& $10$ & $1$ & $1;0,2$ \\
$1$ & $4$ & $0,2,4$ & $1,2,3,4;1,3,5$ \\
& $8$ & $0,2$ & $1,2;1,3$ \\
& $12$ & $0$ & $1$ \\
$2$ & $6$ & $1,3$ & $1,2,3;0,2,4$ \\
& $10$ & $1$ & $1;0,2$ \\
$3$ & $8$ & $0,2$ & $1,2;1,3$ \\
& $12$ & $0$ & $1$ \\
$4$ & $10$ & $1$ & $1;0,2$ \\
$5$ & $12$ & $0$ & $1$ \\
\hline
\end{tabular}
\end{center}
\end{table}
\ed